\documentclass[12pt,preprint]{aastex}
\usepackage{fullpage}
\usepackage[utf8x]{inputenc}
\usepackage{epsf,amsfonts,amsmath,amssymb}
\usepackage{graphicx}
\usepackage{times}
\usepackage{natbib}
\usepackage{ulem}
\usepackage{epsfig}
\usepackage{mathrsfs}
\usepackage[usenames,dvipsnames]{color}

\def\G{\Gamma}

\definecolor{DarkBlue}{rgb}{0,0.08,0.45}

\begin{document}
\title{The emission mechanism in magnetically dominated GRB outflows}
\author{Paz Beniamini\altaffilmark{1a} \& Tsvi Piran\altaffilmark{1b}}
\altaffiltext{1}{Racah Institute for Physics, The Hebrew University, Jerusalem, 91904, Israel}
\email{(a) paz.beniamini@mail.huji.ac.il; (b) tsvi.piran@mail.huji.ac.il}
\begin{abstract}
We consider the conditions within a Gamma-Ray Burst (GRB)
emission region that is Poynting flux dominated. 
Due to the enormous magnetic energy density, relativistic electrons will cool in such a region
extremely rapidly via synchrotron.
As there is no known mechanism that can compete with synchrotron it must be
the source of the prompt sub-MeV emission.
This sets strong limits on the size and Lorentz factor of the outflow. 
Furthermore, synchrotron cooling is too efficient. It overproduces optical and X-ray as compared with the observations.
This overproduction of low energy emission can be avoided if the electrons are re-accelerated many times ($\gtrsim 5\times 10^4$)
during each pulse (or are continuously heated) or if they escape the emitting region before cooling down.
We explore the limitations of both models practically ruling out the later and demonstrating that the former
requires two different acceleration mechanisms as well as an extremely large  magnetic energy to Baryonic energy ratio.
To be viable, any GRB model based on an emission region that is Poynting flux dominated must demonstrate how these
conditions are met. We conclude that if GRB jets are launched magnetically dominated they must dissipate somehow most
of their magnetic energy before they reach the emission region.
\end{abstract}

\section{Introduction}
\label{Int}
It is generally accepted that Gamma Ray Bursts (GRBs) are powered by ultra relativistic jets (
with a Lorentz factor $\Gamma\geqslant100$) carrying
large isotropic equivalent luminosities ($L_{iso}\approx10^{53}$erg/sec) which are dissipated at large distances ($r\approx10^{13}-10^{17}$cm) from the central engine.
Whether the composition of the jets is: Baryonic  \citep{Shemi(1990)} or magnetically (Poynting flux)
 dominated 
\citep{Usov(1992),Thompson(1994),Meszaros(1997),Lyutikov(2003)} is one of the major puzzles concerning GRBs.

The thermal pressure at the base of an AGN jet is insufficient to support a bayronic outflow and therefore,  blazars' jets must be magnetically dominated. 
However, the strong IC component observed in blazers shows that this magnetic energy must have dissipated efficiently before the emission regions and the latter are baryonic dominated
\citep[see e.g.][]{GhiselliniTavecchio09}.
The situation for GRBs could be very different as with smaller size engines the thermal pressure in the base could be sufficient to drive baryonic outflows. However, 
modeling of some GRBs' engines based on accretion disks suggest that the Poynting flux jet power is much
stronger than the thermal driven outflow that derived from neutrino annihilation \citep[see e.g. ][]{Kawanaka(2013)}.
Furthermore, the fact that blazars jets are magnetically driven, suggests that this might also be the case for GRBs. On the other hand, the fact that unlike blazers GRBs
don't have a strong high energy IC component suggests strong magnetic fields within their emission region.

The most direct clues on the nature of the jets arise from the observed prompt sub-MeV spectrum. As such, deciphering the nature of GRB jets is closely related to another
major puzzle concerning GRBs - what is the radiation mechanism that produces the prompt emission. The situation is somewhat inconclusive.
A strong upper limit on the thermal component  \citep[e.g. GRB 080916C][]{Zhang(2009)} suggests a Poynting flux dominated outflow. On the other hand
possible detection of $5-10\%$ of the total energy in a thermal component 
\citep[e.g. GRBs: 100724B, 110721A, 120323A][]{RydePeer(2009),Ryde(2010),Guiriec(2011),Guiriec(2013),Axelsson(2012)} which is not expected in a Poynting flux outflow, indicate a baryonic outflow.

The overall non thermal spectrum has led to the suggestion that the prompt emission is produced by synchrotron mechanism \citep{Katz(1994),Rees(1994),Sari(1996),Sari(1998)}.
This was supported by the reasonable association of the afterglow emission with synchrotron radiation. 
However, observations of many bursts showing lower energy spectral slopes steeper than the ``synchrotron line of death" \citep{Crider(1997), Preece(1998), Preece(2002)}
have led to the suggestion that the prompt emission must be produced by another mechanism.

However, a refined time dependent spectral analysis in several bright bursts \citep{Guiriec(2012)} has
suggested that the situation might be much more complicated and
GRB spectra may be better fitted with a multi component model instead of the classical ``Band function". In the multi component model,
the spectrum is composed of a superposition of a Band function peaking at few hundred keV and carrying most of the energy,
a black body component at a few tens of keV with $5-10\%$ of the energy and in some cases an additional power law component extending up to a few
hundred MeV and carrying up to $40\%$ of the total energy. In these fits, the lower energy spectral slope of the Band function becomes softer
compared with the Band only fits, and is consistent with slow cooling synchrotron (i.e. the synchrotron ``line of death" problem is removed). 
The direct detection of a weak photospheric component in those GRBs, suggests that the ``thermal signature" 
is sub-dominant and it is not the main radiation source in operation during the prompt.
These later results support synchrotron emission as the main radiation mechanism while requiring additional mechanisms to produce the other components.

In this paper, we explore these two strongly connected issues:
the composition of GRB jets and the prompt radiation mechanism. 
Clearly, a Poynting flux dominated jet must dissipate 
a significant fraction of its magnetic energy and produce 
relativistic electrons that would emit efficiently the radiation at the emission zone.
We consider, first, a simple one zone model for the Poyinting flux jet.
Initially (the moment dissipation starts), $\epsilon_B\sim1$, in such scenario. After a fraction $\epsilon_{diss}$ of the magnetic energy
is dissipated, $\epsilon_{B, after}=1-\epsilon_{diss}$ remains as a magnetic energy.
A fraction of the dissipated energy goes into electrons and a fraction of that is radiated away:
$\epsilon_{rad} <\epsilon_e<\epsilon_{diss}$.  \citet{Panaitescu(2002), Granot(2006)} and \citet{Fan(2006)} show that the radiative
efficiency of GRBs is of order unity:  $\epsilon_{rad}\sim 0.2-0.8$. Therefore 
$\epsilon_{diss}\gtrsim 0.2-0.8$ and the ratio of magnetic to electron energy density
at the emitting region is $\epsilon_{B,after}/\epsilon_e<(1-\epsilon_{diss})/\epsilon_{diss} \sim 0.25-4$.
Depending on the time it takes for the magnetic energy dissipation to take place there are two possibilities.
If the dissipation is very rapid (compared to synchrotron cooling),
then the actual radiation process takes place in a low magnetic field region
and the radiation process ``is unaware" of the original Poynting flux dominated nature of the outflow. 
This is similar to the situation in AGNs, where the magnetic energy within the emission region is sub-dominant.
On the other hand, if the dissipation is slower, then the electrons effectively emit most of their energy in a region where the magnetic fields are still relatively high.
We focus on this latter possibility in this paper, and we find constraints that should be imposed on such a scenario. These constraints have to be satisfied by any
radiation process within a Poynting flux dominated emission region in order for it to be viable.

As relativistic electrons are essential to produce the observed prompt emission (regardless of the 
specific emission mechanism) then 
due to the high magnetic energy density, a significant synchrotron signal (not necessarily in the sub-MeV band) must
be produced as well. 
We compare the expected synchrotron fluxes with upper limits on prompt observations at the optical, X-ray and GeV bands,
and show that in order for the synchrotron flux to be consistent with observations, the gamma-ray process should be unrealistically
efficient. Thus, in a magnetically dominated emission region, synchrotron emission
must be the main emission mechanism during the prompt phase. This sets numerous limits on the conditions, such as the
radius and Lorentz factor of the emitting regions. 

Additionally, as the synchrotron is extremely efficient, the electrons will cool rapidly and in addition to producing a
$\nu^{-1/2}$ low energy spectral slope they will over produce optical and X-ray. This is inconsistent with 
observational limits within these energy bands. Thus, this low energy fast cooling tail must be avoided. 
In the last sections of this work we consider a few ways to avoid this problem. 
For example  it has been suggested that this inconsistency may be alleviated if
electrons are re-accelerated many times during a single pulse however this may lead, in turn, to other problems. We also consider three simple two-zone models in which the
magnetic field is inhomogenous. These toy-models sereve as demonstrating the intrinsic limitations of such configurations and highlight issues that must be addressed
by any model of this kind.

This paper is organized as follows. In \S \ref{GenApp} we discuss the basic concepts and the parameter phase space of the model.
In \S \ref{synsig} we discuss the general characteristics of synchrotron emission, focusing
on the typical frequencies and the cooling timescale. In \S \ref{generalconstraints},
we find the constraints on the parameter space from observed limits on the flux in different bands. 
We discuss the implications of the results on the synchrotron 
emission mechanism, including several inhomgenous toy models in \S \ref{synmodel} and we conclude and summarize in \S \ref{conclusions}.

\section{General Considerations}
\label{GenApp}

A magnetically dominated outflow has to dissipate, producing relativistic electrons, in order to generate the observed emission.
An important question is where does this dissipation take place and how efficient is it.
While it is not clear how the dissipation occurs it is clear that within AGN jets 
the outflow that is initially Poynting flux dominated is dissipated efficiently before the emitting region which is baryonic or pair dominated.
It is possible that this might be the case also within GRB jets.
In this case, examination of the resulting spectrum won't reveal the initial nature of the outflow.

A second possibility that we examine here is that a significant fraction of the total energy is still magnetic within the region where the prompt sub-MeV GRB
emission is produced. The following discussion is general and we 
don't specify the emission mechanism producing the prompt gamma-ray signal.
Notice, that we are not assuming anything about the global structure of the jet, only that the
emitting region is magnetically dominated.

Due to the large magnetic energy density, a significant synchrotron signal (not necessarily in the sub-Mev)
is expected. We explore this ``synchrotron signature'' and compare it with observational limits
in various energy bands.

The emitting region associated with the production of a single pulse can be described by 6 parameters (see \cite{Beniamini(2013)}).
These can be chosen as:
the co-moving magnetic field strength, $B'$, the number of relativistic emitters, $N_e$,
the ratio of magnetic to electrons' energy in the jet, $\epsilon$,
the bulk Lorentz factor of the source with respect to the GRB host galaxy, $\G$, the typical electrons'
Lorentz factor in the source frame, $\gamma_m$, and the ratio between the shell crossing time
and the angular timescale, $k$.
The primes denote quantities in the co-moving frame whereas un-primed quantities reflect 
quantities measured in the lab frame.
In principle, a full description should also list $\theta$, the jet opening angle.
However, the source can be treated as spherically symmetric as long as $\G^{-1}<\theta$ which
holds during the prompt phase, where $\G \geqslant 100$.
We therefore do not solve for $\theta$ and use isotropic equivalent quantities throughout the paper.

An upper limit on the emitting radius is given by the variability time scale,
$t_p\geqslant t_{ang}\equiv R/(2 c \G^2)$.
We define a dimensionless parameter $k$ such that $kR/2\G^2$ is the shell's width:
\begin{equation}
\label{t_cross}
t_{cross} \equiv \frac{kR}{2 c \G^2}=kt_{ang}. 
\end{equation}
With this definition, the pulse duration is:
\begin{equation}
\label{t_p}
t_{p} \equiv \frac{(k+1)R}{2 c \G^2}. 
\end{equation}
Clearly, for $k\leqslant1$ the pulse width is determined by angular spreading.

The relativistic electrons are assumed to have a power-law distribution of Lorentz factors:
\begin{equation}
\frac{dN_e}{d\gamma}=C (\frac{\gamma}{\gamma_m})^{-p}, 
\end{equation}
(where C is a normalization constant) which holds for $\gamma_m<\gamma<\gamma_{Max}$. 
The typical Lorentz factor of the electrons, $\gamma_m$ is determined by the 
total internal energy of the electrons:
\begin{equation}
\label{electronsE}
 E_e'=\frac{p-1}{p-2}\gamma_m N_e m_e c^2\equiv\epsilon_e E_{tot}',
\end{equation}
where $N_e$ is the number of relativistic electrons, $E_{tot}'$ is the energy of the flow in the co-moving frame
and the ratio of the total energy in the relativistic electrons to the total internal energy is denoted by $\epsilon_e$.
It is important to stress that in case the synchrotron cooling time is shorter than the dynamical time,
$\epsilon_e$ does not reflect the instantaneous energy ratio of the relativistic electrons to the total energy,
which could be much smaller.
Furthermore, for models invoking re-acceleration or continuous heating (that are further discussed in \S \ref{synmodel}),
Eq. \ref{electronsE} should be multiplied by the number of acceleration processes.
We expect the power-law index $p$ to be of the order $p\sim 2.5$
\citep{Achterberg(2001), Bednarz(1998), Gallant(1999)}.
Indeed this value agrees well with observations of both the GRB prompt and afterglow phases \citep{Sari(1997),Panaitescu(2001)}. 
If $\gamma_{Max}$ arises  due to Synchrotron
losses at the energy where the acceleration time equals to the energy loss time \citep{de Jager(1996)}
then:
\begin{equation}
\label{synmax}
\gamma_{Max}=4 \times10^7 f B'^{-1/2},
\end{equation}
where $f$ is a numerical constant of order unity which encompasses
the details of the acceleration process and $B'$ is the co-moving magnetic field.
In a shock-acceleration scenario this depends on the amount of time the particle spends in the
downstream and upstream regions \citep{Piran(2010), BDK(2011)}.

In a single zone model the magnetic field is assumed to be constant over
the entire emitting region (we explore deviations from this assumption in \S \ref{synmodel}).
The magnetic energy is:
\begin{equation}
\label{energymag}
E_B'=\frac{B'^2}{8 \pi} 4\pi R^2 \frac{kR}{\G}\equiv\epsilon_B E_{tot}',
\end{equation}
where $kR/\G$ is the thickness of the shell in the co-moving frame
and $\epsilon_B$ is the fraction of magnetic to total energy.
By definition in a magnetic dominated outflow (that we consider here) $\epsilon_B \approx 1$.
As mentioned above, it is possible that initially magnetically dominated jets are dissipated efficiently within
the emitting region, and electrons effectively emit most of their energy in a region where the magnetic fields are still
relatively high, with $\epsilon_B\lesssim1$ but larger than say, 0.1. 
We denote those as ``marginally magnetically dominated emission regions".
In order to take into account this possibility
we later re-introduce the dependence on $\epsilon_B$
when presenting the results in \S \ref{synsig} and \S \ref{generalconstraints}.

We define $\epsilon_{\gamma}$ as the efficiency of conversion between total and (isotropic equivalent) radiated energy in gamma rays.
The energy of the observed sub-MeV flux, is huge and already highly constraining astrophysical models.
In addition, observations of afterglows, show that the energy released in the afterglow is at most that
of the prompt \citep{Panaitescu(2002), Granot(2006), Fan(2006)}.
The implication is that a significant amount of the kinetic energy is released in the prompt phase and the efficiency
of the gamma ray emitting process should be high.
Here we choose a canonical value of $\epsilon_{\gamma}=0.1$.
Assuming that the gamma ray emission arises from relativistic electrons, the efficiency is
limited by the amount of energy that passes at some point through the electrons: $\epsilon_{\gamma}<\epsilon_e$.
The requirement $\epsilon_{\gamma}=0.1$ leads to $\epsilon_e>0.1$. The narrow range of $\epsilon_e$, together
with the fact that most relevant parameters depend only weakly on its value, mean that in exploring the parameter space it
is possible to assume a constant $\epsilon_e=0.1$ with no loss of generality.

$\G$ is highly constrained by the opacity to photon-photon pair production
\citep{Fenimore(1993),Piran(1995),Woods(1995),Sari(1999),Lithwick(2001),Nakar(2005),Zou(2010)}
to within the range: $50\leq\G\leq3000$. 
Here we choose, for illustrative purposes, three canonical values: $\G=50,300$ and $1000$.

Two main observations define the relevant parameters 
of the prompt emission process. The duration of a typical pulse and its isotropic equivalent energy. 
Again for illustrative purposes we take fiducial values.
We consider a typical pulse with and isotropic equivalent energy of $E_{tot}=5 \times 10^{52}$erg, and a duration, $t_p=0.5$sec (both in the host frame).
This reduces the number of free parameters from six to two, which we choose as: ($\gamma_m,R$).
Six further constraints described in \S \ref{generalconstraints} further limit the allowed region within this parameter space.

\section{The synchrotron signature}
\label{synsig}
The synchrotron emission is characterized by two frequencies. $\nu_m$,
the synchrotron frequency for the typical energy electron, 
and the cooling frequency, $\nu_c$, the frequency at which electrons cool via synchrotron on the pulse time-scale.
Notice, that in a magnetically-dominated region, IC losses are essentially sub-dominant compared with Synchrotron losses.
The former frequency is given by: 
\begin{equation}
\label{synch1}
\nu_{m}=\G \gamma _m ^2 \frac{q_eB'}{2 \pi m_e c }.
\end{equation}
The later satisfies:
\begin{equation}
\label{coolfreq}
 \nu_c=\frac{18\pi q_e m_e c}{\sigma_T^2 B'^3 \G t_a^2(1+Y)^2},
\end{equation}
where $t_a$ the dynamical time (for which $t_a=t_p$) or the time available for cooling between two consecutive acceleration episodes \citep{Kumar(2008)} whichever is shorter,
and Y is the Compton parameter (that may be ignored for the canonical case of interest here, i.e. $\epsilon_B \approx 1$).
The synchrotron energy flux,
$\nu F_{\nu}$, peaks at $max[\nu_c,\nu_m]$ \citep{Sari(1998)}.

\begin{figure} [h]
\centering
\epsscale{0.6}
\plotone{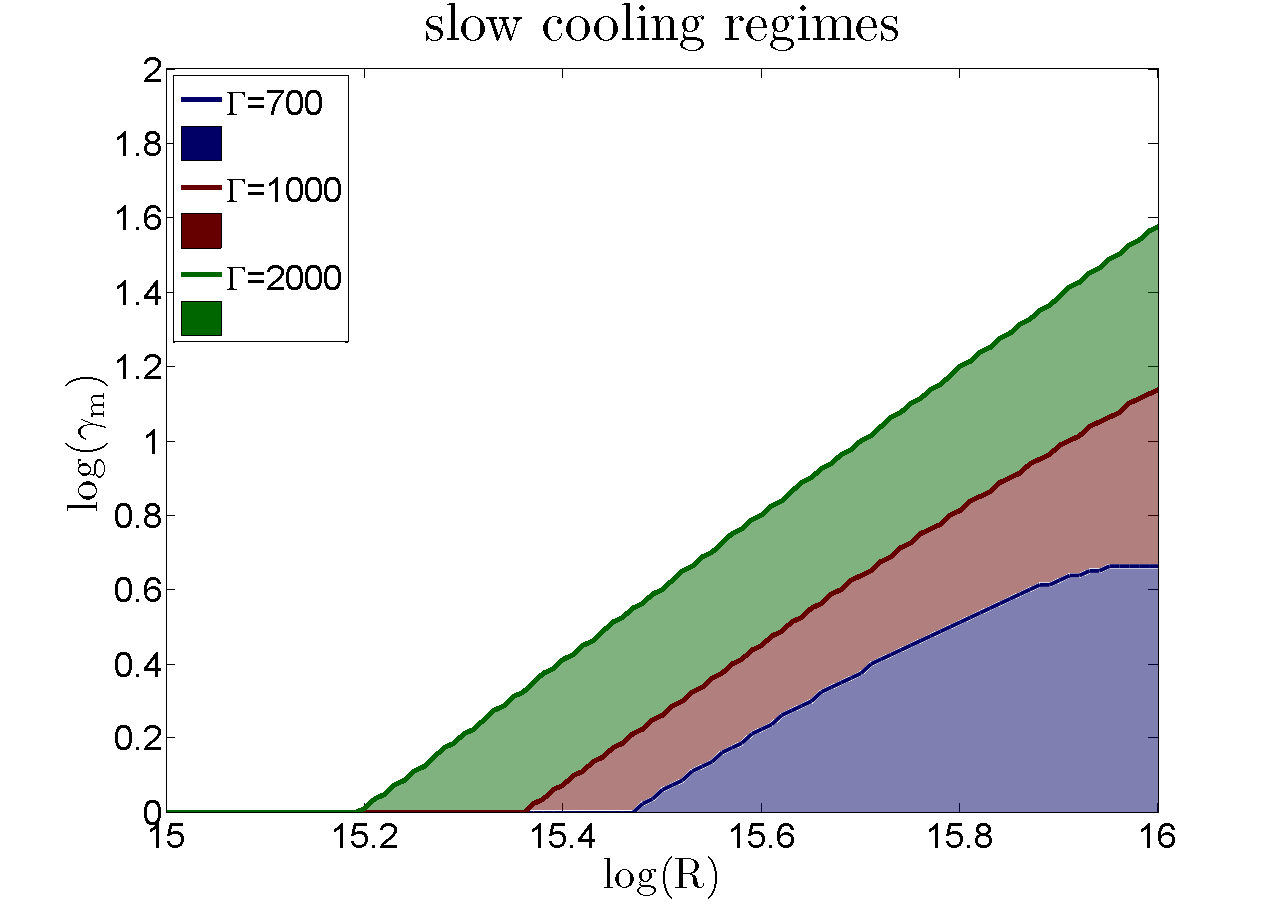}
\caption
{\small Within the colored areas the electrons are slow cooling. Blue, Red and Green correspond to $\Gamma=700, 1000, 2000$ accordingly.
(For values of $\Gamma\lesssim 600$, the synchrotron emitting electrons are always fast cooling).
Unless the bulk Lorentz factor is very high, the electrons are most likely fast cooling.}

\label{fig:slowcooling}
\end{figure}

Assuming one-shot acceleration, $t_a=t_p$. The synchrotron frequency for the typical 
energy electrons is larger than the cooling frequency whenever:
\begin{equation}
\label{coolm}
\gamma_m>0.46  k (k+1)^{-3} \G_{2.5}^{5} E_{tot,52.5}^{-1} t_{p,-0.3}^2 \epsilon_B^{-1}(1+(\frac{\epsilon_B}{\epsilon_e})^{1/2}) ,
\end{equation}
where we use  the notation: $q_x = q/10^x$ in c.g.s. units here and elsewhere in the text.
Fig. \ref{fig:slowcooling} depicts the regions in the parameter space where the electrons are slow cooling.
The fast cooling regime becomes larger for smaller radii (larger values of $k$).
For $\G=50,300$ the synchrotron is in the fast cooling regime independent of $\gamma_m,R$ and $\epsilon_e$.
For $\G=1000$ the synchrotron radiating electrons will be fast cooling, unless: $\gamma_m\lesssim 190 k (k+1)^{-3}E_{tot,52.5}^{-1} t_{p,-0.3}^2
\epsilon_B^{-1}(1+(\frac{\epsilon_B}{\epsilon_e})^{1/2})$
(or equivalently $R_{15}\gtrsim10^{2}\gamma_{m,3.5}^{1/2}E_{tot,52.5}^{1/2} t_{p,-0.3}^{-1}$cm).
Thus, by virtue of the strong magnetic fields of the Poyinting flux dominated jet, quite generally 
$\nu_c\ll\nu_m$. This implies a very short cooling time by synchrotron, as:
$t_{c,syn}/t_p=(\nu_c/\nu_m)^{1/2}$ (where $t_{c,syn}$ is the cooling time for the typical electron, see \cite{Beniamini(2013)} for details).
We note that even for $\epsilon_B=0.1$ we still reach the same qualitative results (i.e. the electrons are fast cooling in most of the parameter space).
In this case, depending on $\epsilon_e$, SSC losses may no longer be neglected, but still  $\nu_c\ll\nu_m$. 

The synchrotron cooling time and the typical frequency may be affected by self absorption.
If self absorption takes place ($\nu_{SSA}>\nu_m$, where $\nu_{SSA}$ is the frequency below which synchrotron photons are self absorbed),
the synchrotron spectra peaks at $\nu_{SSA}$. In this case, the total luminosity radiated by synchrotron is reduced and the typical cooling
time increases by the ratio of the unabsorbed to absorbed synchrotron luminosities.
By equating the flux expected from synchrotron to that of a black body with a temperature $k_BT \approx \gamma_m m_ec^2$, one finds that the synchrotron self absorption frequency 
will be:
\begin{equation}
\label{eq:ssa}
\left\{
  \begin{array}{l l}
    \nu_{SSA}= 10^{14} E_{tot,52.5}^{3/10} \Gamma_{2.5}^{2/5}\gamma_{m,3.5}^{-4/5} R_{15}^{-3/5}t_{p,-0.3}^{-3/10} \mbox{Hz}, & \quad \nu_m<\nu_{SSA}\\
    \nu_{SSA}= 10^{15}A_P E_{tot,52.5}^{\frac{2+p}{10}} \Gamma_{2.5}^{2/5}\gamma_{m,3.5}^{\frac{2p-6}{5}} R_{15}^{-\frac{2+p}{5}}t_{p,-0.3}^{-\frac{2+p}{10}} \mbox{Hz}, & \quad 
    \nu_{SSA}<\nu_m\\
  \end{array} \right.
\end{equation}
Where $A_p$ is a numerical function of $p$ which is of order unity for $p\sim 2.5$ and $\gamma_m$ is taken to be of order $3000$ which
is its value in case synchrotron radiation is the mechanism producing the sub-MeV peak \citep{Beniamini(2013)}.
For sufficiently low values of $\gamma_m$ it is possible 
that because of self absorption
the synchrotron cooling time will increase, but even with this {modification}
the cooling time will be  much shorter than the dynamical one.

\begin{figure} [h]
\centering
\epsscale{0.9}
\plotone{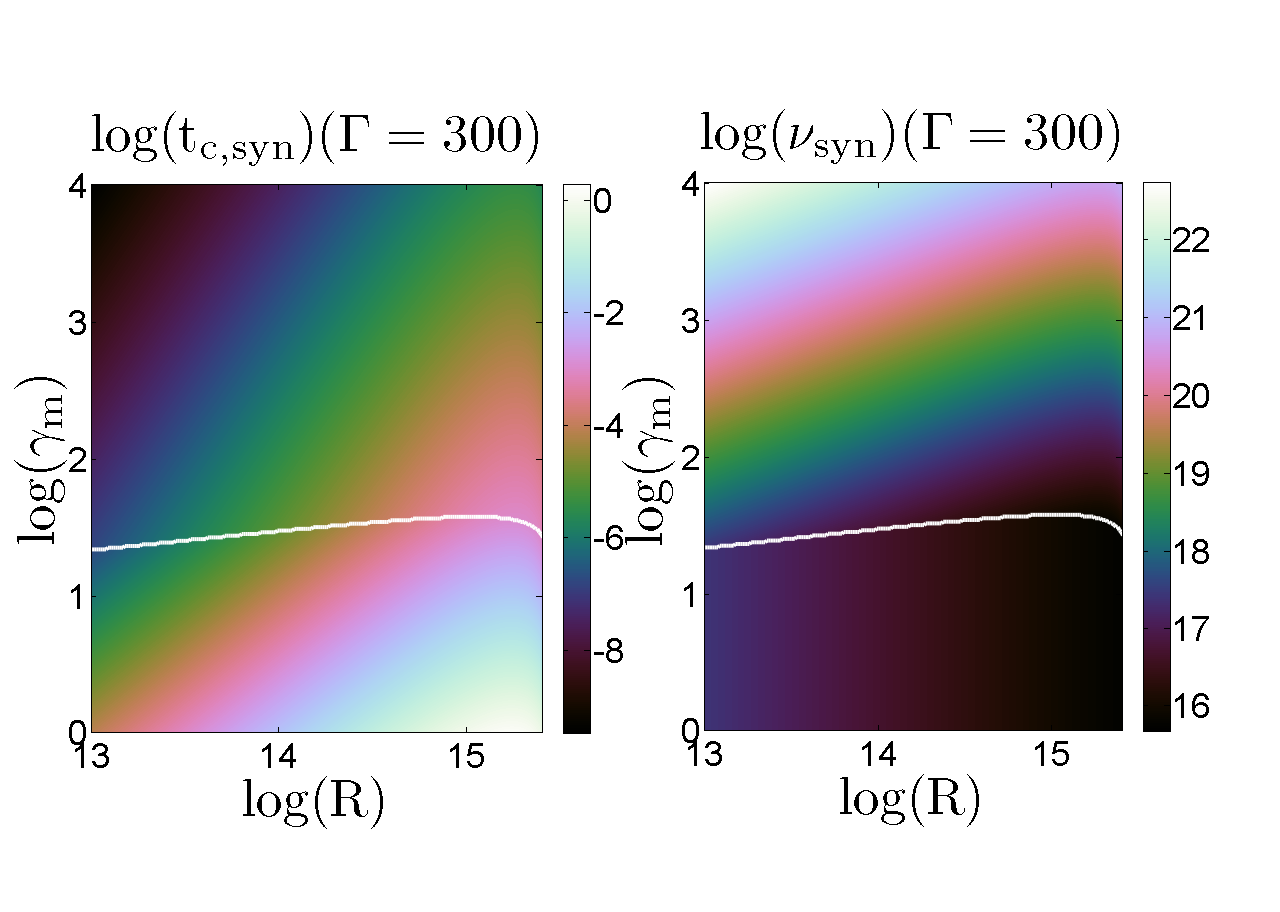}
\caption
{\small Cooling time by synchrotron (left panel) and the synchrotron peak frequency (right panel).
The peak frequency is either $\nu_m$ (above the white line) or $\nu_{SSA}$ (below the line) in cases where the synchrotron signal becomes self absorbed (corresponding to the lower part of the figure).
In the self absorbed regime, the frequency depends slightly on $\epsilon_e$. However, in the interest of clarity, we neglect this dependence in these plots and
assume a constant $\epsilon_e=0.1$ in the entire parameter region. The synchrotron can peak anywhere between EUV and HE gamma-rays and cools very rapidly.}\label{fig:coolsynplusnu}
\end{figure}

We plot the cooling time by synchrotron and the peak frequency of the synchrotron signature in Fig. \ref{fig:coolsynplusnu}
for the ``canonical" case of $\G=300$. The peak synchrotron frequency is either $\nu_m$ (for $\G=300$ the synchrotron is always ``fast cooling") or
$\nu_{SSA}$ in cases where the synchrotron becomes self absorbed.
We find that the synchrotron can peak anywhere between EUV and HE gamma rays, and has a cooling time shorter than $\sim 10^{-1}$sec.
The  short synchrotron cooling time implies that synchrotron is very  efficient  and the electrons will radiate all their energy  within a time scale 
much shorter than a dynamical time \footnote{The fact that the synchrotron
cooling timescale is very short was already mentioned almost 20 years ago by \cite{SariPiran(1995)}, though in a somewhat different context.}.
Any emission  mechanism competing with synchrotron must be even faster in order to 
tap the electrons' energy  before they are cooled by synchrotron.
Even if this mechanism is more efficient than synchrotron (i.e. has a shorter cooling time), the accompanying
synchrotron signal may still overproduce optical and X-ray fluxes compared with observations.
We consider this possibility in the following sections and find limits on the cooling time associated with any emission process producing the
prompt sub-MeV emission in order to sufficiently quench the unobserved synchrotron signature. 
If synchrotron is responsible for the sub-MeV emission, as can be expected in a magnetic dominated outflow,
the electrons must be re-accelerated before cooling to suppress this low-energy emission. In this case 
these limits set the re-accelerating time needed.

\section{General Constraints - limits on the parameter space}
\label{generalconstraints}

We turn now to consider different constraints on the parameter phase space. Some of those constraints are generic while others depends on specific details of the model. 
The non-thermal high energy tail of the spectrum indicates that  the source must be (at least marginally) optically thin for scatterings
\footnote{In photospheric models, the radiation is emitted from the photosphere. Even in this case the optical depth at the emission region cannot be larger than 10}:
\begin{equation}
\label{tau1}
\tau=\frac{N_{e} \sigma_T}{4(k+1) \pi R^2}\lesssim10.
\end{equation}
Note that even for $\tau \geq 1$ (for which the  photons typically encounter more than one relativistic electron before escaping the
emitting region) SSC losses cannot change significantly the synchrotron spectrum as long as the magnetic energy density is comparable or larger
than the electrons' energy density. Specifically, \citep{Daigne(2011)} have shown that both $Y \gg 1$ and strong Klein Nishina suppression 
are needed in order for SSC to significantly affect the synchrotron spectrum. They find that $\epsilon_B \lesssim 10^{-3}$ is needed in order 
to change the typical synchrotron spectrum.

Notice that for a fixed total  electrons' energy  $N_e  \propto \gamma_m^{-1}$. Therefore:
\begin{equation}
\gamma_{m,3.5}^{-1} \epsilon_{e,-1} R_{15}^{-1} \G_{2.5}^{-3}E_{tot,52.5} t_{p,-0.3}^{-1}<10^4.
\end{equation}

If the flow is composed of an electron-proton plasma (as opposed to pairs),
the energy of the protons should be smaller than the magnetic energy,
hence:  
\begin{equation}
\frac{\gamma_m}{\epsilon_e} \frac{m_e}{m_p} \frac{p-1}{p-2}>1,
\end{equation}
where $m_p$ is the proton mass.
As this limit is more model dependent than the previous one, 
we do not use it as a rigid limit on the parameter space.

As mentioned above, SSC is an inevitable byproduct of synchrotron radiation. This emission should also be below the observed limits. A full analysis
of SSC can be found in \citep{Nakar(2009),Bosnjak(2009)}, and a simplified one zone model is described in \cite{Beniamini(2013)}.
For magnetically dominated outflows in the Thomson regime, the relative SSC to synchrotron power is given by $(\epsilon_e/\epsilon_B)^{1/2}$
(in the KN regime it is even smaller).
Therefore, SSC  cooling, is sub-dominant to synchrotron cooling.
The SSC peaks at approximately $\nu_{SSC}\approx\gamma_m^2 \nu_m$. Since the observed limits on the flux are frequency dependent,
it is in principle possible that for some region in the parameter space, some of the observational limits
will be more constraining for IC than for synchrotron.
We have explicitly taken into account these limits from SSC everywhere in the parameter space, but as it turns out they do not further constrain
the allowed parameters. Therefore we disregard the SSC contribution in the rest of the discussion.

\begin{figure}
\centering
\epsscale{0.7}
\plotone{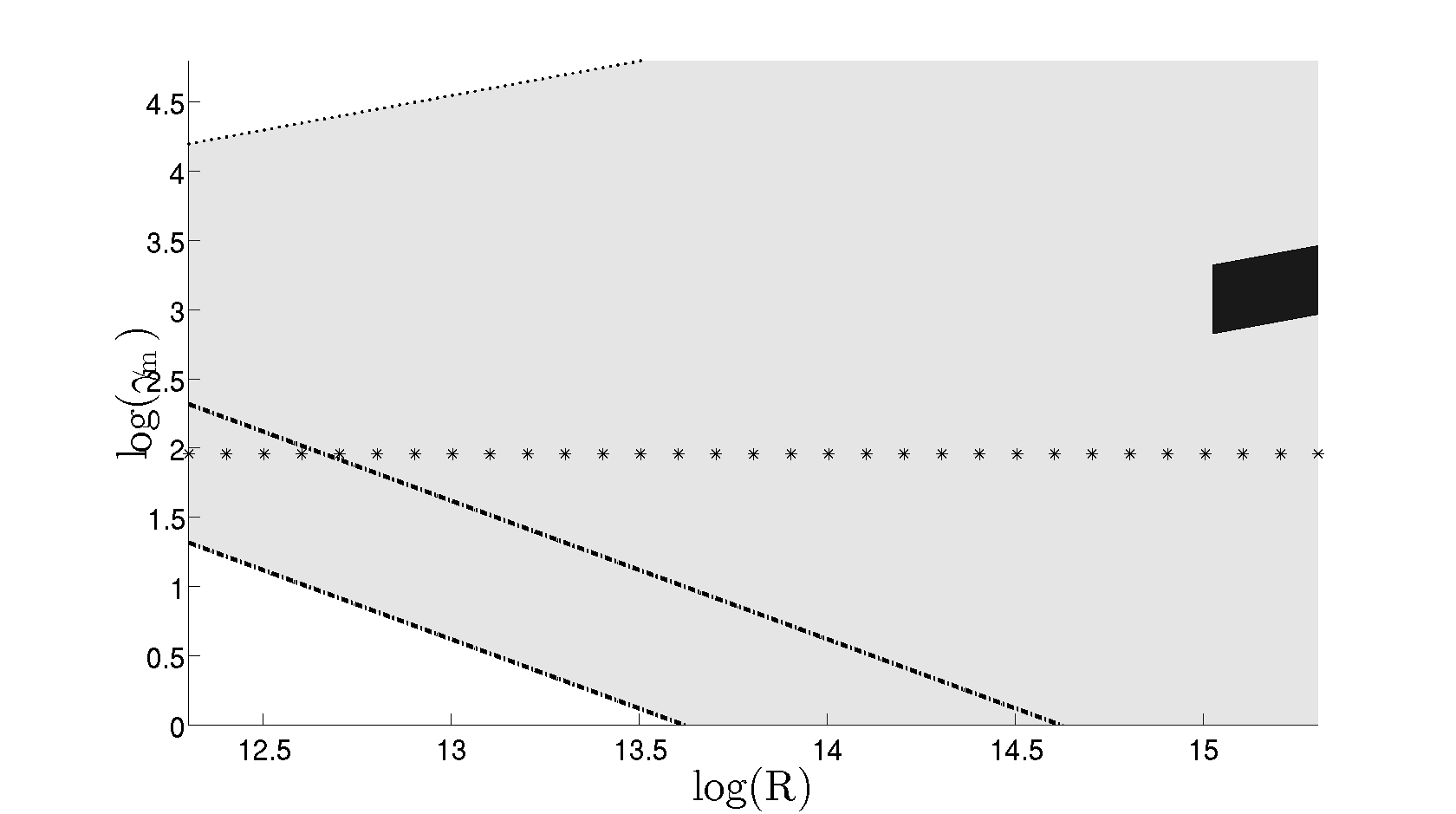}\\

\caption
{\small The  allowed region in the parameter phase space (gray region)  in a magnetically
dominated jet with $\G=300$. The conditions $\tau<1,10$ (dot-dashed lines) and $\gamma_m<\gamma_{max}$ (dotted line) define general limits on the parameter space.
If the flow is composed of a proton-electron plasma, the energy of the protons
must be smaller than the magnetic energy (that is assumed to dominate). This condition rules out the area below the asterisks for such jets.
The black filled region is where synchrotron emission produces the main prompt sub-MeV peak.
}\label{fig:areas}
\end{figure}

The allowed parameter space (for the ``canonical" case of $\Gamma=300$), taking into account the above limits, is shown in Fig. \ref{fig:areas}.
The black area in the figure, is where the synchrotron signal produces the main sub-MeV peak (\cite{Beniamini(2013)}).

\subsection{Constraints on the parameter space from the observed fluxes}
\label{observedlimits}

Observations in the optical, X-ray and GeV bands provide further limits on the gamma ray emitting mechanism.
They show that $t_c$ (the cooling time associated with the  gamma ray emitting
mechanism) has to be significantly smaller than the synchrotron cooling time, $t_{c,syn}$ in order not to overproduce
synchrotron flux as compared with observations in these bands. 
Alternatively, the synchrotron emitting electrons must be re-accelerated on this time scale.
The synchrotron flux in any band is proportional to the ratio of the cooling time of the (undetermined) prompt emission mechanism to the overall cooling times:
$x_t\equiv {t_c}/(t_{c,syn}+t_c)\leqslant1$.  
In a given observed band, one requires $t_c/t_{c,syn}<F_{\nu_{obs,A}}/F_{\nu_{syn,A}}$ where $F_{\nu_{syn,A}}$ is
the synchrotron spectral flux and $F_{\nu_{obs,A}}$ is the observed limit on the spectral flux and A stands for opt, X or GeV. 

\subsubsection{Optical limits}

Prompt optical fluxes (in the V band) span almost five orders of magnitude in flux, from 100$\mu$Jy to 3Jy ($7.5<V<19$) while contemporaneous gamma ray detections
span from 6$\mu$Jy to 4mJy % (14.5<V<21.5)
\citep{Yost(2007)}.
find that for individual bursts, the ratio $F_{\nu_{obs,\gamma}}/F_{\nu_{obs,opt}}$,
of gamma-ray to optical flux is somewhat less variable, and spans from: $2\times 10^{-3} - 1$. 
Using TAROT observations in the optical (R) band, \cite{Klotz(2009)} estimated that $5-20\%$ of GRBs have prompt optical emission at the level of $\sim 10$mJy ($R<14$)
while more than $50\%$ of bursts have prompt optical fainter than $2$mJy ($R=15.5$).
As a typical limit, we take here $F_{\nu_{obs,opt}}=1$mJy at the V-band.

To estimate the synchrotron flux we recall that
there are three relevant orderings of the synchrotron frequencies $\nu_{SSA}$, $\nu_a$,  and $\nu_m$ which cover the majority of cases considered here.
Consider first, $\nu_c<\nu_{SSA}<\nu_{opt}<\nu_m$ which happens whenever:
\begin{equation}
 \gamma_{m,3.5}>0.15 k^{1/8} (k+1)^{5/8} \epsilon_{e,-1}^{1/2} \G_{2.5}^{-1} E_{tot,52.5}^{3/8} t_{p,-0.3}^{-9/8} \nu_{opt,14.7}^{-5/4} x_t^{1/2},
\end{equation}
In this case, the synchrotron optical flux is not self absorbed: $F_{\nu_{syn,opt}}=F_{\nu_m} ({\nu_m}/{\nu_{opt}})^{1/2}$.
As a result, the limit from observations in the optical band becomes highly constraining in this regime:
\begin{equation}
\label{gmopt}
\gamma_{m,3.5}>10^3\epsilon_{e,-1}\G_{2.5}k^{1/4} (k+1)^{-3/4}E_{tot,52.5}^{3/4} t_{p,-0.3}^{-1/4} \nu_{opt,14.7}^{-1/2}F_{\nu_{opt,1mJy}}^{-1}x_t,
\end{equation}
where $F_{\nu_{opt,1mJy}} \equiv F_{\nu_{obs,opt}}/1mJy$, $\epsilon_{e,-1}=\epsilon_e/0.1$. 
Since $F_{\nu_{syn,opt}}\propto\gamma_m^{-1}$, $\gamma_m$ has to be large in order for the synchrotron to
peak at  energies much beyond the optical band and the  optical synchrotron flux will be sufficiently low.
In order to satisfy this limit the cooling time of
the gamma ray emitting mechanism should be significantly shorter than the synchrotron cooling time i.e. a low value of $x_t$.

For intermediate values of $\gamma_m$:
\begin{equation}
0.03 k^{\frac{3}{14}} (k\!+\!1)^{\frac{-5}{14}} \epsilon_{e,-1}^{\frac{1}{7}} \G_{2.5}^{\frac{3}{7}} E_{tot,52.5}^{\frac{-1}{7}} t_{p,-0.3}^{\frac{3}{14}}x_t^{\frac{1}{7}}\!<\! \gamma_{m,3.5}\!<\!
0.15 k^{\frac{1}{8}} (k\!+\!1)^{\frac{5}{8}} \epsilon_{e,-1}^{\frac{1}{2}} \G_{2.5}^{-1} E_{tot,52.5}^{\frac{3}{8}} t_{p,-0.3}^{\frac{-9}{8}} \nu_{opt,14.7}^{\frac{-5}{4}} x_t^{\frac{1}{2}},
\end{equation}
the optical spectrum is suppressed due to self-absorption (i.e. $\nu_{opt}<\nu_{SSA}$).
In the thermal part of the spectrum, the flux scales as $F_{\nu} \propto R^2 T$ (where $T\propto \gamma_m$ is the temperature of the electrons).
Thus, for sufficiently low values of $\gamma_m$, the temperature and hence optical flux will be  below the observed limit,
independently of the total energy of the outflow, $E_{tot}$.
The corresponding limit on $\gamma_m$ and $R$ is:
\begin{equation}
\label{SAlimitgm}
R_{15}^2\gamma_{m,3.5}<3\times 10^{-5} \G_{2.5} \nu_{opt,14.7}^{-2} F_{\nu_{opt,1mJy}} x_t^{-1}.
\end{equation}

Finally, for:
\begin{equation}
\gamma_{m,3.5}<0.03 k^{3/14} (k+1)^{-5/14} \epsilon_{e,-1}^{1/7} \G_{2.5}^{3/7} E_{tot,52.5}^{-1/7} t_{p,-0.3}^{3/14}x_t^{1/7},
\end{equation}
the typical synchrotron frequency, $\nu_m$, drops below the self absorption frequency.
Here, the order of frequencies is $\nu_c<\nu_{opt},\nu_m<\nu_{SSA}$. As discussed above, this does not change the limits on the optical flux,
which are given directly by the radius of emission and temperature of the electrons.
Nonetheless, as the synchrotron peak is self absorbed, the total energy released by synchrotron is reduced in this case,
and this causes an increase in the synchrotron cooling time compared with the non-absorbed case (as seen below
the white line in Fig. \ref{fig:coolsynplusnu}).

\begin{figure}
\centering
\epsscale{0.7}
\plotone{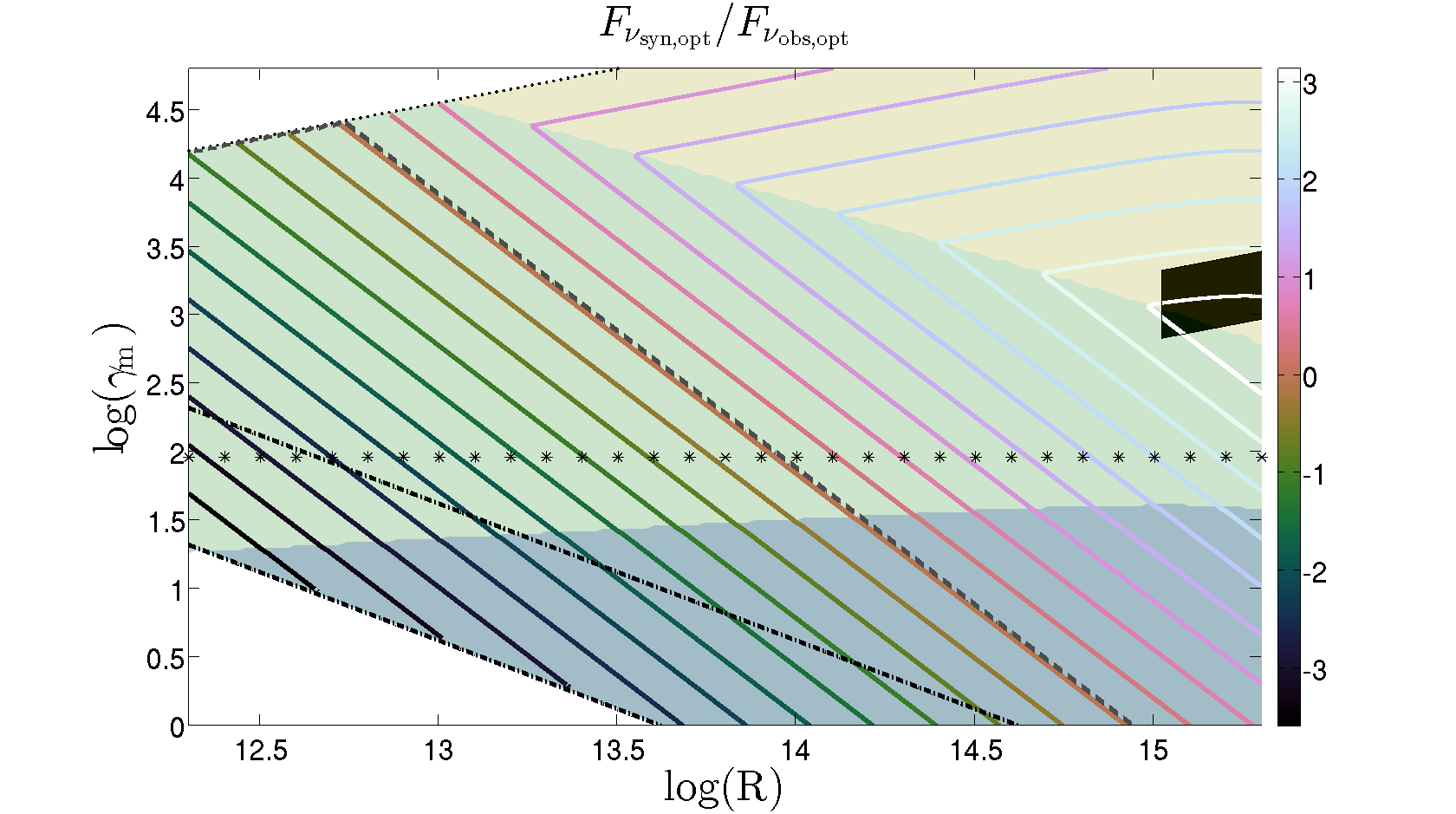}\\
\caption
{\small Same setup as in Fig. \ref{fig:areas}.
The background colors in these regimes, signify the various possible orderings of frequencies.
From top to bottom these are:
$\nu_m,\nu_x>\nu_{opt}>\nu_{SSA}>\nu_c$ (no self absorption at the optical band; yellow area), $\nu_m,\nu_x>\nu_{SSA}>\nu_{opt}>\nu_c$
(self absorption at the optical band but not at the peak; green area)
and $\nu_x>\nu_{SSA}>\nu_m,\nu_{opt}$ (both the peak and the optical band of the synchrotron are self absorbed; blue area).
The ratio of synchrotron optical flux to the observed optical flux is depicted by contour lines.
To the right of the gray dashed line, this ratio is larger than one. In this region, the gamma ray process should
cool considerably faster than synchrotron in order to reduce the synchrotron flux below the observed limit.
}\label{fig:optical}
\end{figure}

Fig. \ref{fig:optical} depicts $F_{\nu_{syn,opt}}/F_{\nu_{obs,opt}}$ on top of the filled regions which signify the three
orderings of frequencies that were discussed above. Large $R$ and $\gamma_m$ lead to $F_{\nu_{syn,opt}}/F_{\nu_{obs,opt}}\gg1$
and subsequently yield strong upper limits on the required cooling time of the gamma ray emission mechanism.

\subsubsection{X-ray limits}

In some GRBs XRT,  the X-ray telescope on board the Swift satellite,  observed the bursts   during the prompt phase
(either in cases where there was a precursor to the main burst or when the burst was very long).
From these bursts one can obtain limits on the prompt X-ray flux of GRBs. These turn out to be of the order of
0.5-10mJy ($5\times 10^{-27}-10^{-25}\mbox{erg }\mbox{sec}^{-1} \mbox{Hz}^{-1}\mbox{cm}^{-2}$) \citep{Burrows(2005),Campana(2006),Krimm(2006),Moretti(2006),Page(2006),Godet(2007)}
at the median energy of XRT, which is $\sim2$keV.
In all these cases the prompt X-ray lightcurve was seen to track the gamma-rays.
Here we pick a canonical limit of $F_{\nu_{obs,X-rays}}=1$mJy at 2keV.

\begin{figure}
\centering
\epsscale{0.7}
\plotone{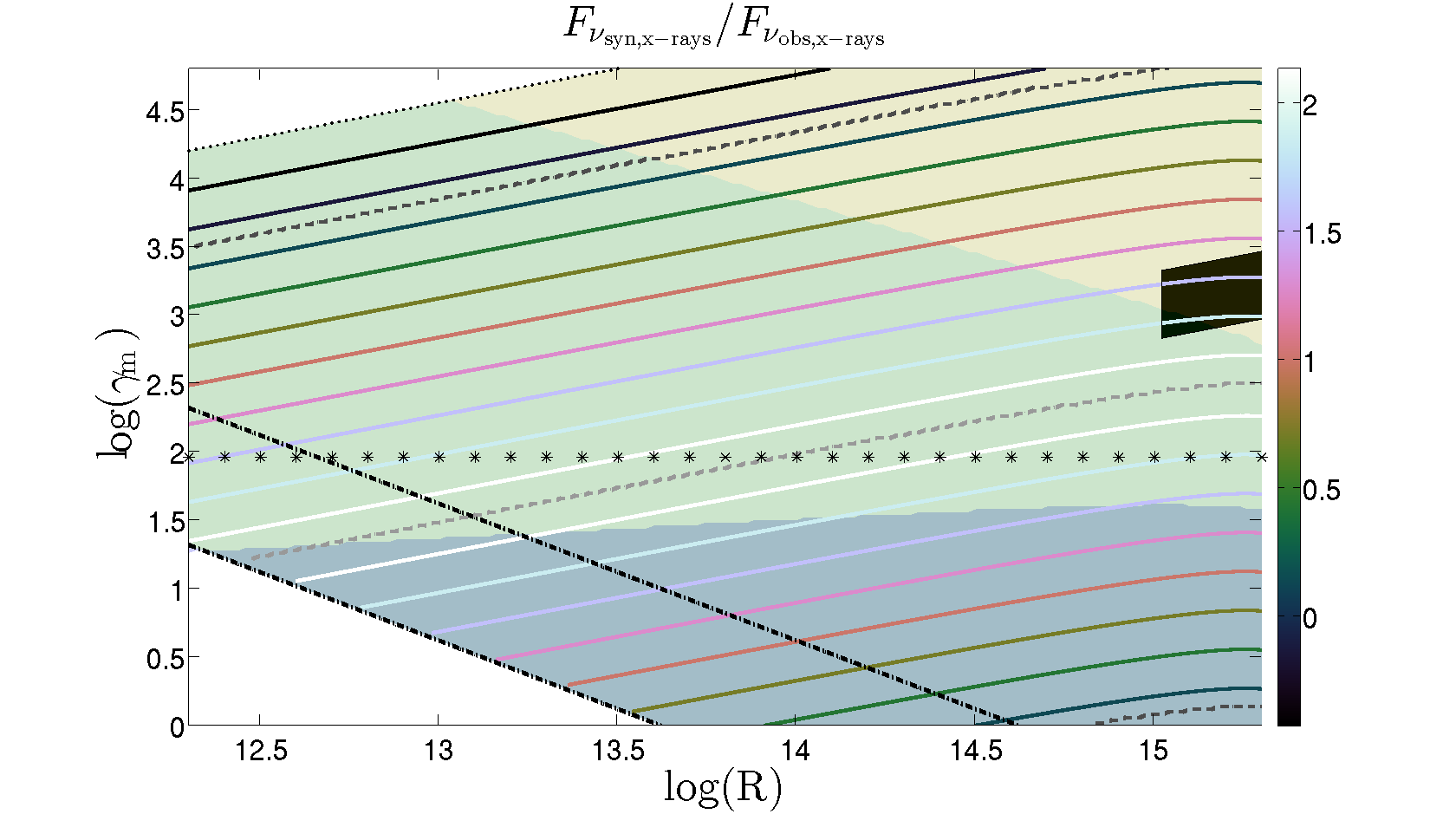}\\
\caption
{\small Background colors and lines are the same as in Figs. \ref{fig:areas},\ref{fig:optical}.
The ratio of synchrotron X-ray flux to the observed X-ray flux is depicted by contour lines. Between the dashed
gray lines, the ratio is larger than one, requiring the gamma ray process to be more efficient than synchrotron (i.e. have a short cooling time).
The lighter dashed gray line in the middle of the figure is $\nu_m=\nu_x$ (below it the X-ray intercepts the high energy part of the synchrotron
spectrum and above it the low energy part).
}\label{fig:xray}
\end{figure}

By virtue of Eq. \ref{eq:ssa} the X-ray band will not be self-absorbed.
Therefore, there are two relevant orderings of the frequencies with regard to the X-ray band. If $\nu_m>\nu_x$, the X-ray band falls in the
low energy part of the synchrotron spectrum. In this case, as $\gamma_m$ increases, the synchrotron peak frequency is shifted to higher energies, until it is eventually
so high, that the extrapolation of the flux to the X-ray band is below the observed limits. In essence, this is the same kind of limit as given by Eq.\ref{gmopt}
{but now it is applied to} the X-ray instead of to the optical band.
If $\nu_m<\nu_x$, the X-ray band falls in the high energy part of the synchrotron spectrum. In this case, as $\gamma_m$
decreases, the synchrotron peak frequency decreases and the synchrotron flux in the X-ray band decreases as well.
These two cases can be written as:
\begin{equation}
\left\{
  \begin{array}{l l}
    \gamma_{m,3.5}>40\epsilon_{e,-1} \G_{2.5} k^{1/4} (k+1)^{-3/4} E_{tot,52.5}^{3/4} t_{p,-0.3}^{-1/4}\nu_{X,17.7}^{-1/2}F_{\nu_{X,1mJy}}^{-1}x_t \mbox{cm}, & \quad \nu_x<\nu_m\\
    \gamma_{m,3.5}<8\times10^{-4} \epsilon_{e,-1}^{-1} \G_{2.5} k^{1/4} (k+1)^{-3/4} E_{tot,52.5}^{-5/4} t_{p,-0.3}^{7/4} \nu_{X,17.7}^{3/2}F_{\nu_{X,1mJy}}x_t^{-1} \mbox{cm}, & \quad 
    \nu_m<\nu_x\\
  \end{array} \right.
\end{equation}
where $\nu_{X,17.7}$ is the XRT median frequency, and $F_{\nu_{X,1mJy}} \equiv F_{\nu_{obs,X-rays}}/1mJy$.
Fig. \ref{fig:xray} depicts $F_{\nu_{syn,X-ray}}/F_{\nu_{obs,X-ray}}$. Unless $x_t\ll1$, 
either very high or very low values of $\gamma_m$
(corresponding to $\nu_m\gg\nu_x$ or vice versa) are required in order not to overproduce X-ray radiation.

\subsubsection{GeV limits}
Another observational limit arises from observations of $\sim 20\mbox{MeV} - 30\mbox{GeV}$ photons from GRBs with LAT, the HE telescope on board the Fermi mission.
From LAT observations it is found that the fluence in the LAT band during the prompt phase is less than 0.1
of the sub-MeV fluence \citep{Guetta(2011),Beniamini(2011), Ackermann(2012)}. Most bursts are consistent with an extrapolation of
the sub-MeV flux. However, about a 1/3 of the bursts have smaller fluxes at these high energies than expected by extrapolation of the peak flux.
Additionally, bursts with LAT detections show a correlation between GBM (sub-MeV) and LAT fluxes.

These limits correspond to $\nu_{syn,LAT} F_{\nu_{syn,LAT}} > 10^{-7}\mbox{erg sec}^{-1}\mbox{cm}^{-2}$.
As $\gamma_m$ increases, the synchrotron peak frequency increases 
and the fraction of the flux that falls within the LAT band increases.
Therefore this corresponds to an upper limit on $\gamma_m$ (see Fig. \ref{fig:LAT}):
\begin{equation}
\gamma_{m,3.5}<3 \epsilon_{e,-1}^{-1} \G_{2.5} k^{1/4} (k+1)^{-3/4} E_{tot,52.5}^{-5/4} t_{p,-0.3}^{7/4}.
\end{equation}
As the low flux in the LAT band could be due to a cut-off in the synchrotron spectrum (e.g. due to pair creation opacity),
this is a somewhat weaker constraint than those given by the optical and X-ray fluxes. 
These  are relevant, of course, only  for models where a pair creation cutoff is 
expected at higher energies. 

\subsection{Results - Limits on the Electrons' Cooling Time}
\label{cooltimes}
\begin{figure}
\centering
\epsscale{0.7}
\plotone{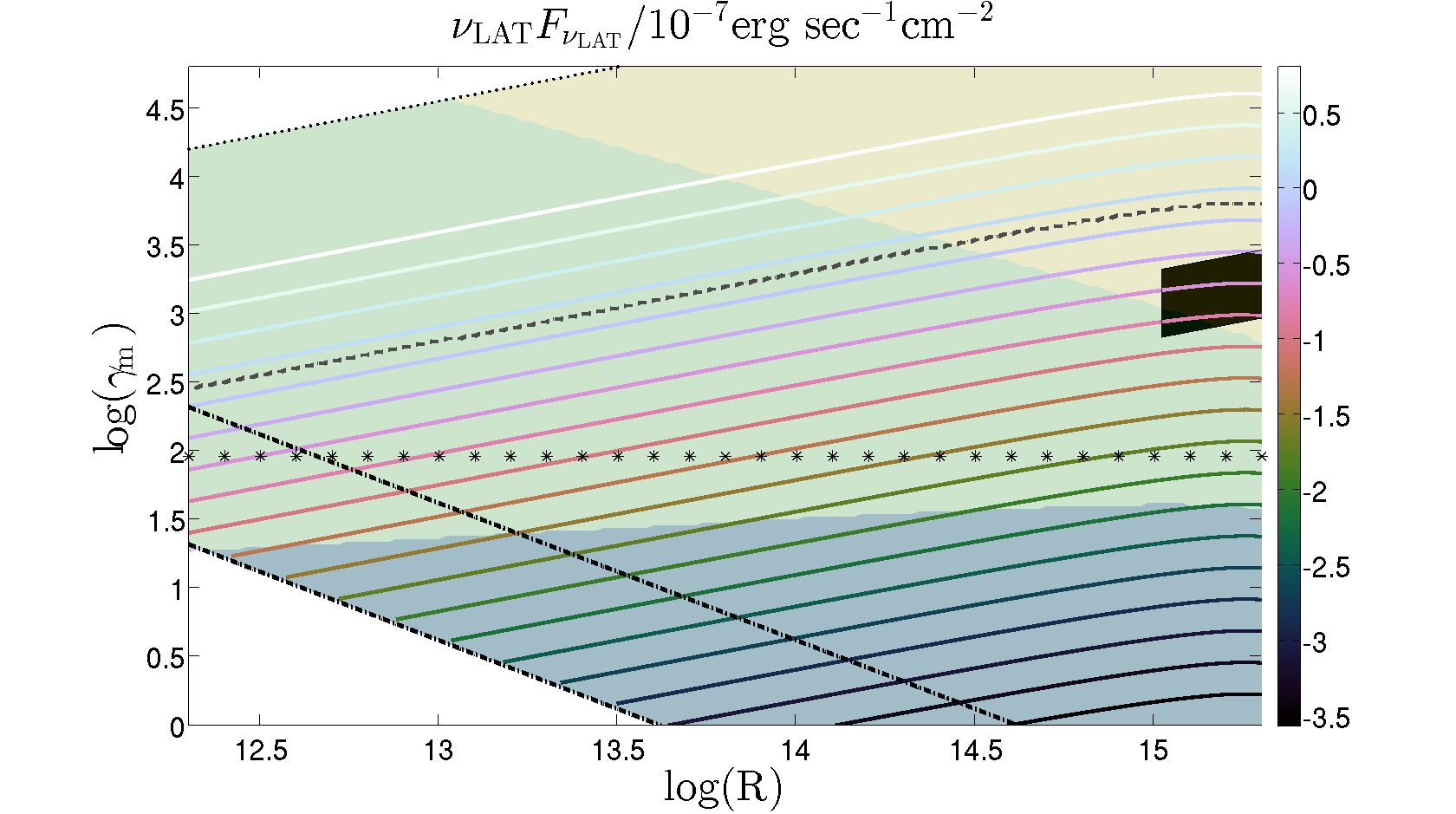}\\
\caption
{\small Background colors and lines are the same as in Figs. \ref{fig:areas},\ref{fig:optical}.
The ratio of synchrotron $\nu F_{\nu}$ in the LAT band to the observed limits in the same band is depicted by contour lines.
Above the dashed gray line, the ratio is larger than one, requiring the gamma ray process to be more efficient than synchrotron.
}\label{fig:LAT}
\end{figure}

\begin{figure}
\centering
\epsscale{0.7}
\plotone{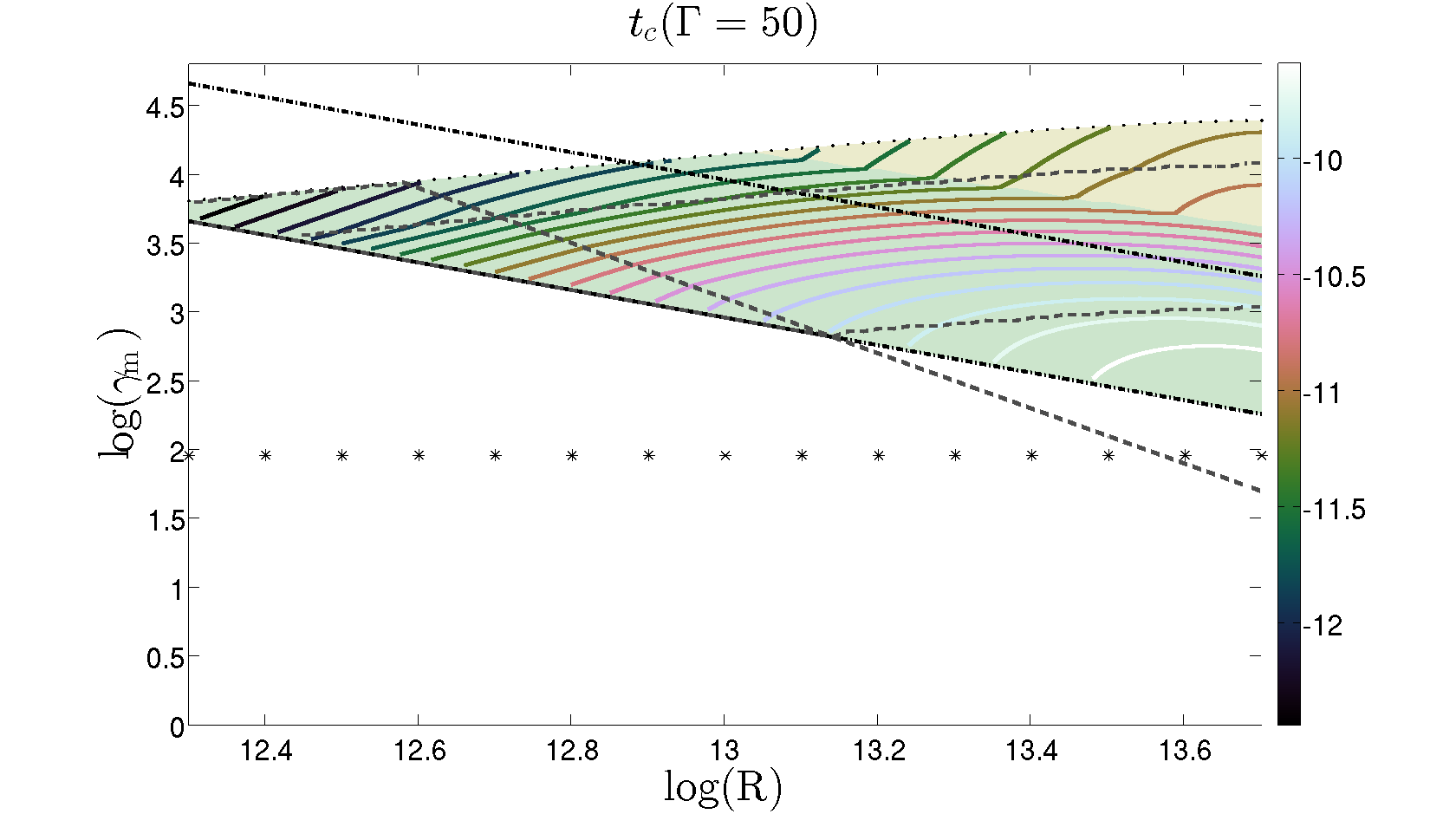}\\
\caption
{\small Same setup as Figs. \ref{fig:areas}, \ref{fig:optical} for $\G=50$.
In addition, contour lines depict the maximum allowed value for the cooling timescale ascosiated with the gamma-ray
producing mechanism ($t_c$) in order not to overproduce synchrotron emission.
$t_c$ should be shorter than $t_{c,syn}$ (the synchrotron cooling timescale) in order for the gamma ray 
process to be able to tap a significant amount of the electrons' energy.
The conditions $F_{\nu_{syn,opt}} < 1\mbox{mJy}$, $F_{\nu_{syn,X-ray}} < 1\mbox{mJy}$
and $\nu_{syn,LAT} F_{\nu_{syn,LAT}} < 10^{-7}\mbox{erg sec}^{-1}\mbox{cm}^{-2}$ further constrain $t_c$.
Beyond the corresponding lines, $t_c$ should be significantly shorter than $t_{c,syn}$ in order for the synchrotron not to
overproduce optical, X-ray or GeV radiation which are unobserved.
}\label{fig:G50}
\end{figure}

\begin{figure}
\centering
\epsscale{0.7}
\plotone{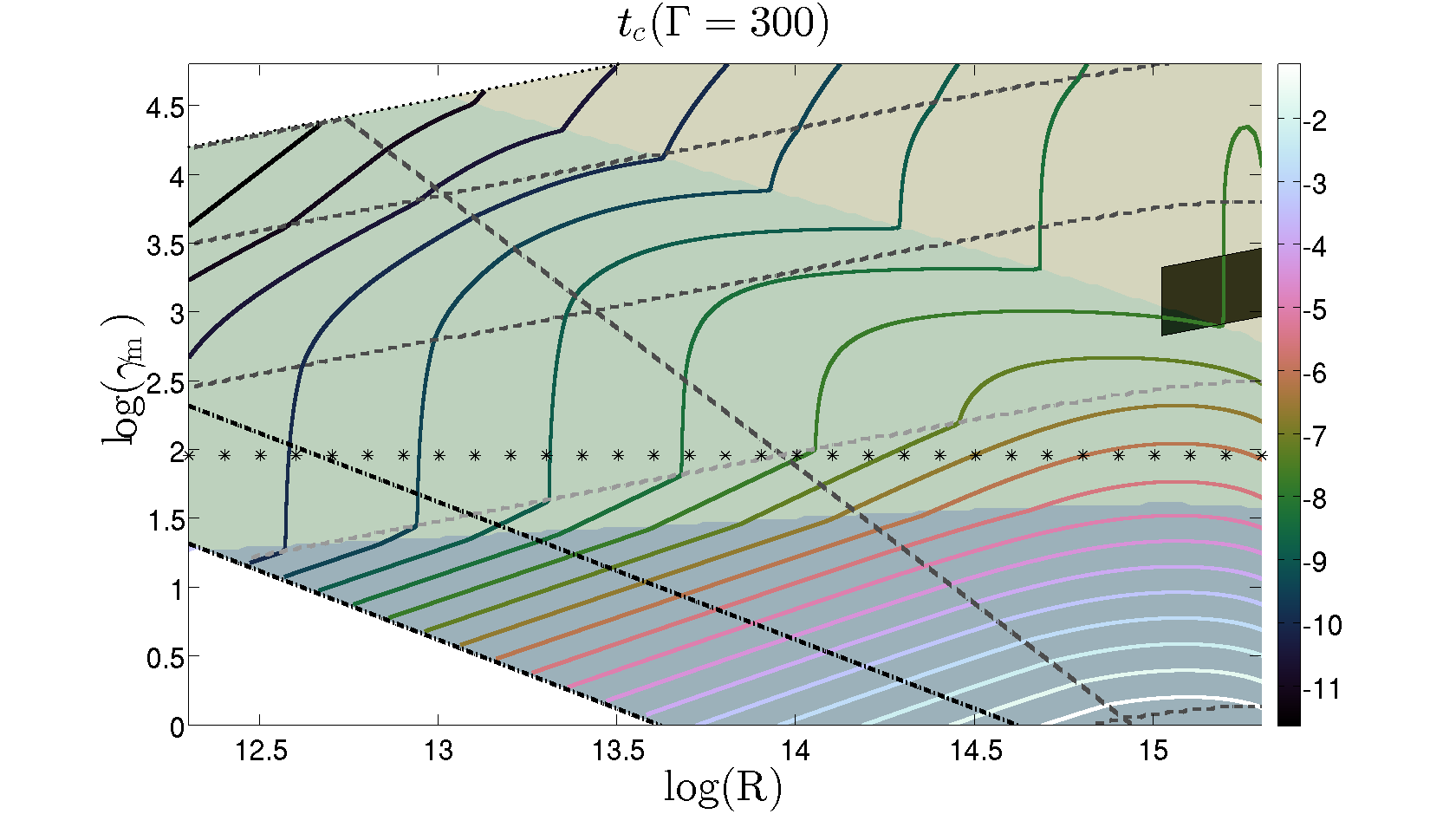}\\
\caption
{\small Same as in Fig. \ref{fig:G50} for $\G=300$.}\label{fig:G300}
\end{figure}

\begin{figure} [h]
\centering
\epsscale{0.7}
\plotone{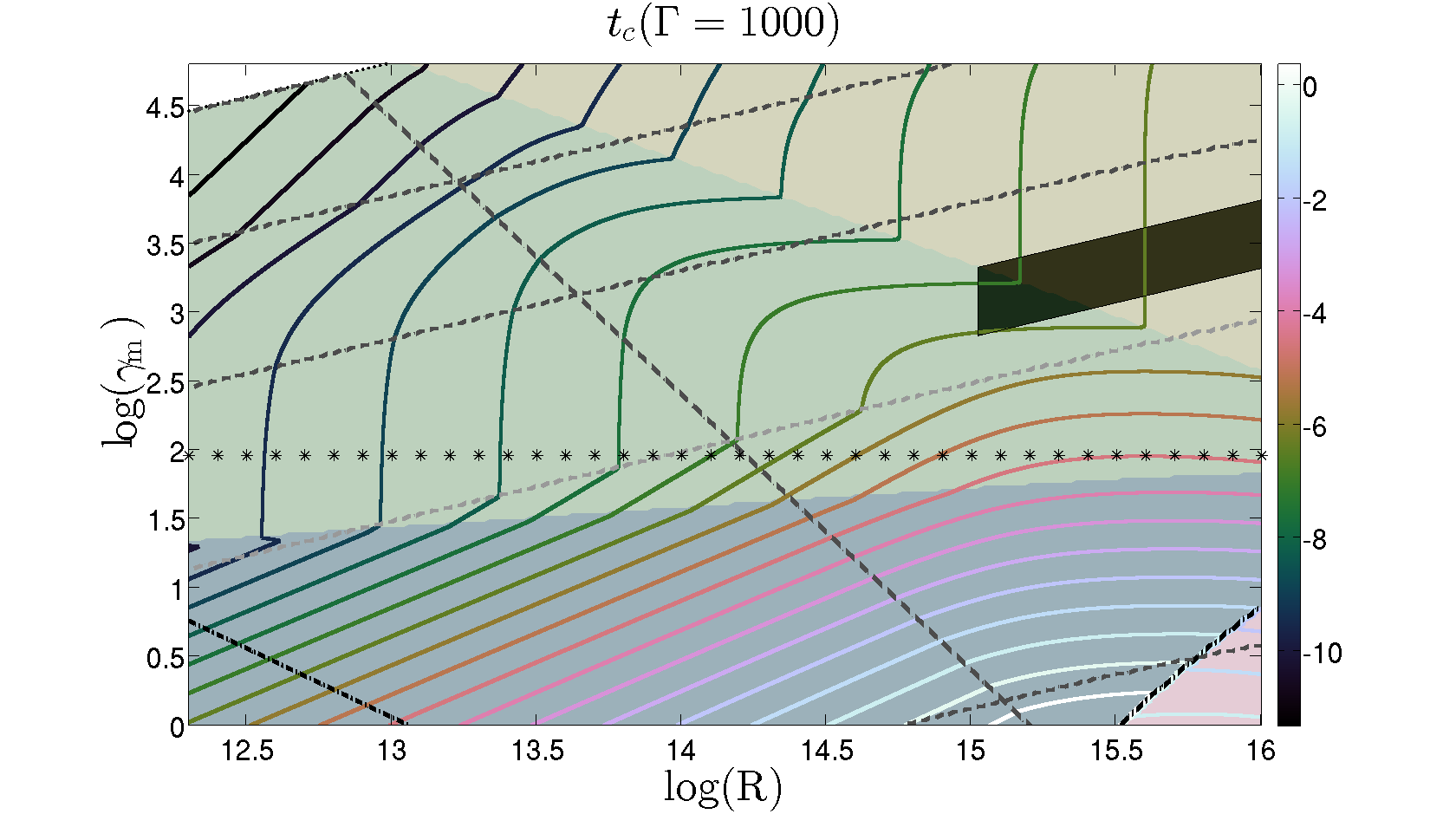}
\caption
{\small Same as in Fig. \ref{fig:G50} for $\G=1000$. 
The red area at the bottom of the $\G=1000$ plot (which did not appear previously),
signifies where the synchrotron emitting electrons become
slow cooling (same as in Fig. \ref{fig:slowcooling}).}\label{fig:G1000}
\end{figure}

\begin{figure} [h]
\centering
\epsscale{0.7}
\plotone{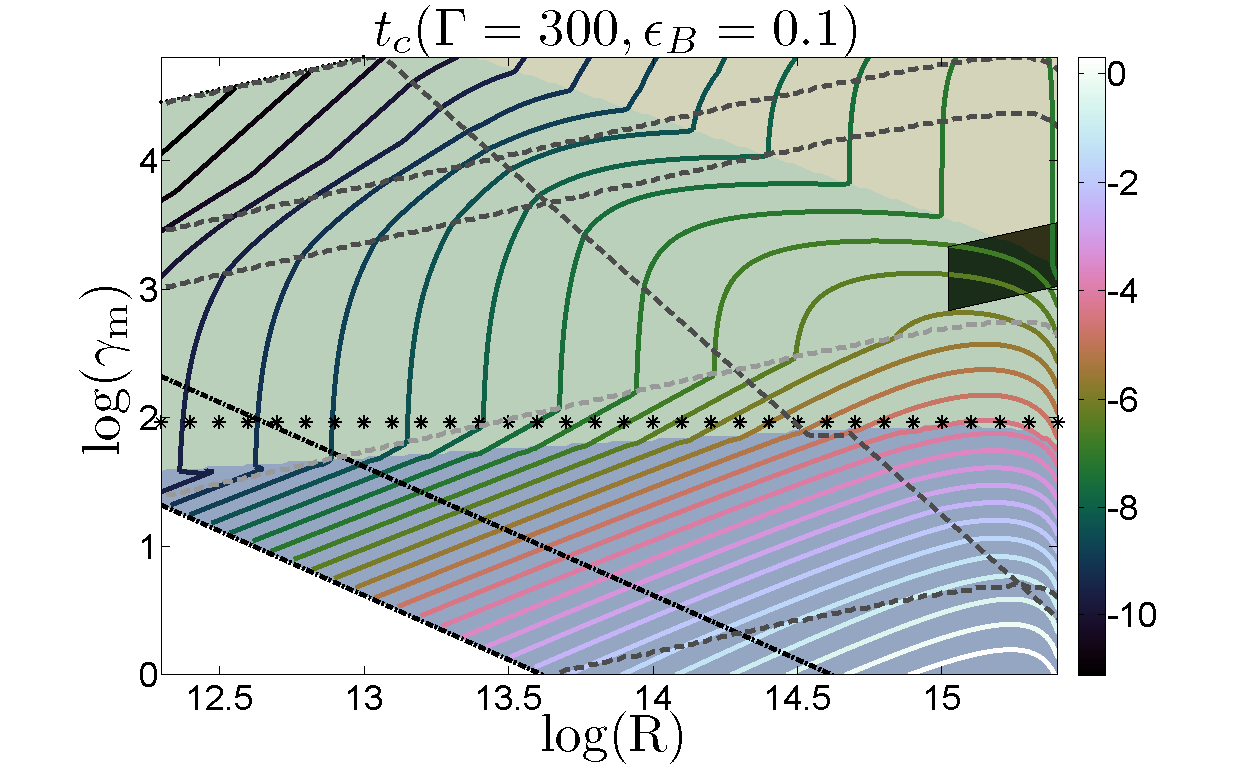}
\caption
{\small Same as in Fig. \ref{fig:G300} for $\epsilon_B=0.1$.
The synchrotron cooling time is at most increased by a factor of 10 comapred to the $\epsilon_B=0.1$ case, while
in most of the parameter space it remains very similar to the previous possibility.}\label{fig:G300B01}
\end{figure}

As shown in Figs. \ref{fig:optical}, \ref{fig:xray}, \ref{fig:LAT}, observations in the optical, X-ray and GeV bands
provide even stronger limits on the efficiency of the gamma ray emitting mechanism.
$t_c$ has to be significantly shorter than the synchrotron cooling time, $t_{c,syn}$ in order not to overproduce
synchrotron flux as compared with observations in these bands.
The combined limits from the optical, X-ray and GeV on the cooling time are shown in 
Figs. \ref{fig:G50}, \ref{fig:G300}, \ref{fig:G1000} for $\G=50,300,1000$ respectively.
Almost everywhere in the parameter space the gamma-ray emitting process has to be extremely efficient in order
to avoid excess synchrotron flux in one of the observed bands. The upper limits on $t_c$ typically lie in the range $10^{-11}-10^{-2}$sec.
The upper limits on $t_c$ become lower with decreasing radius of emission as well as with increasing $\epsilon_e$ and $\gamma_m$.
Finally, we explore the dependence of the results on $\epsilon_B$ in order to address  ``marginally magnetically dominated emission regions".
Fig. \ref{fig:G300B01} depicts the combined limits on the cooling time for $\G=300$ (as in Fig
\ref{fig:G300}) but for $\epsilon_B=0.1$. The upper limit on $t_c$ increases at most by a factor of 10 and 
in most of the parameter space it remains very similar to its value for $\epsilon_B=1$.
We conclude that even for marginally magnetic jets, the limit on the cooling time
is very short in the vast majority of the parameter space.

\section{Implications for the Synchrotron model}
\label{synmodel}
The discussion so far has been general and
could be applied to any emission process. It explores the constraints
that synchrotron radiation imposes when the outflow is Poynting flux dominated.
We turn now to explore the implication of the above results
when synchrotron itself is the source of the prompt emission.
This is in fact the most obvious situation when a magnetic field dominates the energy density in the emitting region,
\citep{Meszaros(1997),Piran(2005)}.
In a previous work \citep{Beniamini(2013)} we have examined the 
general observational constraints on the synchrotron emission of the prompt sub-MeV radiation, for a general magnetic field.
The resulting parameter space found in that work (in the limit of large $\epsilon_B$) is reproduced here in the region of the 
parameter space marked as  black areas in Figs. \ref{fig:optical}, \ref{fig:xray}, \ref{fig:LAT}.

Given the constraints on the optical and X-ray bands considered in \S \ref{observedlimits}, the simple one zone synchrotron model for the prompt gamma rays is ruled out
for a Poynting flux dominated flow. With such strong magnetic fields, the electrons overproduce both optical and X-ray emission. 
This inconsistency is related but not equivalent to the familiar result that the gamma-ray spectrum below the observed sub-MeV peak is, quite often, inconsistent with 
$N_\nu \propto \nu^{-3/2}$ predicted for fast-cooling synchrotron \citep{Cohen(1997),Crider(1997), Preece(1998), Preece(2002)} and sometimes even with the slow cooling synchrotron,
the so called ``synchrotron line of death problem".
Clearly, any model that decreases sufficiently fast below the sub-MeV peak 
will be consistent with the optical and X-ray fluxes, which were obtained assuming that the electrons are fast cooling all the way to the lower energy bands.
However,  it is possible that refined analysis of the gamma ray spectra will alleviate the problems concerning the gamma-ray spectrum below the sub-Mev peak.
For example, recent time resolved spectral analysis \citep{Guiriec(2012)} suggests that GRBs are better fitted
with a three component model: the Band function, a black body and a power law.
Most of the energy is carried out by the Band component.
However, the Band component alone is insufficient to fit the entire spectrum.
This more complicated fit will change, of course, the relevant low energy spectral index.
Given this uncertainty we ignore here this issue of the low energy spectral index and we focus 
on the optical and X-ray excess problems. As we will see shortly the two models  we propose essentially 
resolve this issue as well.

In order that the synchrotron X-ray flux (at 2keV) will be consistent with the observed limits,
the synchrotron cooling frequency must be above the X-ray band.
This will lead to a slope (in $N_\nu$) of $\nu^{-2/3}$ up to $\nu_c$ followed by the regular (steeper) fast cooling slope of $\nu^{-3/2}$ above $\nu_c$.
As a result, the X-ray flux will be lower as compared with the $\nu_c<\nu_x$ case. 
The condition $F_{\nu_{syn,X-rays}}<F_{\nu_{obs,X-rays}}$, implies\footnote{Similar estimates  can be applied to the optical band, but these yield weaker constraints on $\nu_c$}:
\begin{equation}
 (\frac{\nu_{x}}{\nu_c})^{1/3}(\frac{\nu_c}{\nu_m})^{-1/2} F_{\nu_m}<F_{\nu_{obs,X-rays}}
\end{equation}
or:
\begin{equation}
\label{Eq:nuc}
h\nu_c> (h\nu_m)^{3/5} (h\nu_{x})^{2/5}\approx 40 \mbox{keV} 
\end{equation}
where the peak of $\nu F_{\nu}$ is at $\nu_m$ for the fast-cooling synchrotron scenario explored here,
and the last equality is for a typical $h\nu_m\sim300$keV (host frame). 

\cite{Beniamini(2013)} have shown that it is possible to achieve such high values of $\nu_c$ within the context of the regular ``one zone" synchrotron model in a non Poynting-flux
dominated emission region. This can happen if 
the emitting radius and $\G$ are sufficiently large.
However, in the magnetically dominated case, the required values of  $R\sim 3 \times 10^{17}$cm and $\G\sim2500$ are unrealistically large.
These are inconsistent with estimates of the deceleration radius for
magnetically dominated jets \citep{Granot(2012)} and with upper limits on $\Gamma$ given by the peak time of the reverse shock and the lack of
afterglow signal between prompt pulses \citep{Sari(1999),Zou(2010)}.
Alternatively, large values of $\nu_c$ can be reached in two ways. Either the electrons are re-accelerated before they cool significantly
(the ``re-acceleration model") or the electrons spend most of their time in regions with low magnetic fields, where they cool less efficiently.
The latter possibility can be achieved in two ways. In both, the magnetic field in the emission zone is very in-homogeneous.
In the ``escape model", electrons emit mainly in small sub-volumes with high magnetic field but they escape these regions before cooling down significantly.
The electrons practically don't emit in most of the volume where the magnetic field is lower. 
Alternatively, in the ``confinement model", electrons are accelerated and radiate in small sub-volumes with a low magnetic field, where effectively $\epsilon_B$ is very low.
The electron are confined to these regions and they don't escape to the high magnetic field background in which they would have cooled very rapidly.
In the following we discuss three simplified two-zone toy models of these configurations and find what conditions must be satisfied for this type of models to be applicable.

\subsection{Re-acceleration}
\label{reacc}
The simplest variant of the re-acceleration model is if the emitting electrons are continuously re-accelerated while emitting \citep{Ghisellini(1999),Kumar(2008),Fan(2010)}.
A second possibility is to abandon the ``one zone" approximation. In this case the electrons are accelerated sporadically within ``acceleration sites",
while they emit throughout the whole system. Namely, the acceleration sites are immersed in a background ``radiation zone" in which the electrons cool down by synchrotron.
A critical condition is that a typical electron reaches an acceleration site and is re-accelerated before it cools significantly, thus avoiding excess low energy emission. 
We keep the nature of the acceleration sites arbitrary, they can be either magnetic reconnection sites, internal shocks or some other plasma instabilities,
and discuss some general properties of this model.

Eqs. \ref{Eq:nuc} and \ref{coolfreq} imply that, to avoid the low energy emission, the time between accelerations is: 
\begin{equation}
t_a \approx 2\times10^{-5} ~\G_{2.5}^{4}(k+1)^{-9/4}k^{3/4}E_{tot,52.5}^{-3/4} t_{p,-0.3}^{9/4}~\mbox{sec}.
\end{equation}
For canonical parameters, this is $\sim 5\times 10^4$ shorter than the pulse's duration.
Namely, to avoid excessive cooling each particle has to be accelerated $5\times 10^4$ times during a single pulse.
This implies that, compared with the number of electrons needed for the single-shot acceleration, $5\times 10^4$
times fewer electrons are needed to produce the same flux. 
Assuming an electron-proton plasma and one proton per relativistic electron (the later is justified below), this yields a ratio between magnetic and Baryonic
flux of:
\begin{equation}
 \sigma=3\times 10^5 \G_{2.5}^{-3}(k+1)^{3/2}k^{-1/2}E_{tot,52.5}^{1/2} t_{p,-0.3}^{-1/2}.
\end{equation}
This ratio is extremely large and as such it poses another constraint on the system.
This high $\sigma$ parameter will affect the jet dynamics. Specifically, the reverse external shock
associated with such highly magnetized flows is very weak and this could affect the early afterglow signal.

Each acceleration episode has to recreate, (more or less) the
initial energy spectrum of the electrons $dN/d\gamma \propto \gamma^{-p}$ for $\gamma_m<\gamma<\gamma_{max}$ with $\gamma_m\gg1$.
We emphasize that it is important that each re-acceleration episode should produce electrons with Lorentz factors starting from $\gamma_m$ up to at
least $20\gamma_m$, in order to recreate the high energy spectral slope, extending almost 3 orders of magnitude above the peak.
As the time needed to accelerate a particle to a given energy is roughly proportional to this energy this sets a 
crucial limitation on the model.
Before each acceleration, the accelerated electrons are at $\gamma_c$ (where $\gamma_c$ is the Lorentz factor of an electron radiating synchrotron at
$\nu_c$: $\nu _{c}=\G \gamma _c ^2 \frac{q_eB}{2 \pi m_e c }$) and a typical electron is accelerated to $\gamma_m$. 
By virtue of Eq. \ref{Eq:nuc}, $\nu_m\gtrsim10\nu_c$, thus $\gamma_m\gtrsim3 \gamma_c$.
This means that a typical electron should be accelerated only by a factor of 3 in each re-acceleration episode.
This implies immediately that this model must involve two different acceleration processes.
First, an initial acceleration process that accelerates most of the electrons to a large Lorentz factor:
\begin{equation}
\label{gm}
 \gamma_m =5000 \G_{2.5}(k+1)^{-3/4}k^{1/4}E_{tot,52.5}^{-1/4} t_{p,-0.3}^{3/4}.
\end{equation}
Then, a re-acceleration process accelerates an already relativistic electron only modestly. 
Hence the two acceleration processes must be very different. 
A similar situation occurs in the ICMART model \citep{Zhang(2011)}.
This complication is of course an intrinsic drawback of the model. Furthermore, 
the first acceleration process must accelerate all electrons without leaving behind any slow ones
(as opposed to what is seen for example in PIC simulations by \cite{Spitkovsky(2008),Sironi(2011)}).
Otherwise, unless a specific mechanism prevents the re-acceleration sites from accelerating mildly relativistic electrons, these electrons 
will be eventually accelerated and produce spurious emission that will contradict the observations.

Outside the acceleration sites, the electrons emit synchrotron radiation while propagating 
along the field lines. The propagation time between sites should be at most $t_a$ in order to avoid excess soft radiation.
The mean free path between encounters with acceleration sites, $\lambda'$, is simply:
\begin{equation}
\lambda'=c t_a'=\Gamma c t_a=2\times 10^8 \G_{2.5}^{5}(k+1)^{-9/4}k^{3/4}E_{tot,52.5}^{-3/4} t_{p,-0.3}^{3/2} \mbox{cm}
\end{equation}
where primes denote quantities in the comoving frame.
This determines the number of sites ($N_{s}$) in terms of the size of an acceleration site, $l'$:
\begin{equation} 
N_{s}=\frac{4 R^3 k}{\G \lambda' l'^2}. 
\end{equation}
We denote by $d'$ the typical distance between acceleration sites. Using 
\begin{equation}
4\pi R^2 \frac{kR}{2 \Gamma}\approx N_s\frac{4\pi}{3} (\frac{d'}{2})^3 , 
\end{equation}
we obtain:
\begin{equation}
\label{Lout}
d'=(\frac{12k R^3}{\Gamma N_{s}})^{1/3}=(3\lambda' l'^2)^{1/3}\lesssim \lambda'.
\end{equation} 
Eq. \ref{Lout} can also be written as:
\begin{equation}
\label{Lout2}
\frac{l'}{d'}=(\frac{l'}{3\lambda'})^{1/3}.
\end{equation}
The smallest plausible size for an acceleration site is the Larmour radius of the highest energy electrons one wishes to accelerate.
Given that the high energy power law above the peak of the observed Band function extends up to $\sim 500\nu_p$,
there should be electrons with Lorentz factors up to $\gamma\approx20\gamma_m$.
Taking the Larmour radius of these electrons as the minimal size of an acceleration site we obtain:
\begin{equation}
\label{contlimit}
0.03\G_{2.5}^{-1/3} <\frac{l'}{d'}<1.
\end{equation}
where the exact value depends on the acceleration mechanism.
These results suggest that while the acceleration sites could be small they can also approach the 
 continuous limit.
In continuous acceleration, an equilibrium may
be reached between heating and cooling such that electrons of some specific energy remain at a constant energy,
and do not overproduce low energy photons \citep{Kumar(2008)}. In this case, $\nu_c \approx \nu_m$
and the low energy spectral slope problem is resolved as well.

\subsection{The escape model}

Alternatively, the spurious low energy synchrotron emission can be avoided if the electrons escape from the emitting region before they cool down.
In this way the effective cooling frequency $\nu_c$ will be determined by the escape time of the electrons from  the emitting region and not by the hydrodynamic time scale.
As the size of the system is of order $c t_{hyd}$ we cannot expect that the electrons will physically leave the region.
However, if the system is sufficiently inhomogeneous such that in addition to the high magnetic field region in which the electrons cool rapidly there are regions of
low magnetic field in which the electrons are slow cooling, then the electrons can escape the emitting zones before cooling significantly.
We discuss here the possible structure of such a configuration.

We consider a model in which there are two typical magnetic fields. A strong field with a 
(co-moving) magnetic field strength $B_s'$ and a weak field region where the
magnetic field is $B_{out}'$, $\eta_B\equiv B_s'/B_{out}'$. 
The fields are such that
electrons radiate efficiently only while passing through the ``radiation sites" in which the magnetic field is strong. 
As we are considering a Poynting flux dominated system in which the magnetic fields carry most of the energy of the system there are two different regimes. 
In the first the magnetic field energy in the background dominates the energy of the system while in the second the total energy of the radiation sites dominates.

An electron with a Lorentz factor $\gamma$ spends a time $t_{s} (\gamma)$ in the radiation sites.
In case the electron only encounters one site during the pulse duration, this is the escape time from a single site.
Otherwise, it is the total time it spends within different radiation sites.
In order to maintain the given sub-MeV peak flux, emitted by electrons with $\gamma> \gamma_c$, while reducing the synchrotron optical and X-ray fluxes that arise from electrons with 
$\gamma<\gamma_c$, $t_{s} $ must satisfy:
\begin{equation}
\label{defescape}
\left\{
  \begin{array}{l l}
    t_{s} (\gamma)<t_{c,s}(\gamma), & \quad \gamma<\gamma_m/3\\
    t_{s} (\gamma)>t_{c,s}(\gamma), & \quad 
    \gamma>\gamma_m\\
  \end{array} \right.
\end{equation}
where $t_{c,s}$ is the synchrotron cooling time in the radiation sites.
Since $t_c(\gamma) \propto \gamma^{-1}$, the escape mechanism must satisfy
$t_{s} (\gamma) \propto \gamma^{a}$ with $a\geq-1$ (the case of $a=-1$ is considered in detail below).
We define $\gamma_{esc}$ as the Lorentz factor of electrons for which $t_{s} (\gamma)=t_{c,s}(\gamma)$.
For $a\geq-1$, electrons with $\gamma > \gamma_{esc}$, will cool before escaping, whereas
electrons with $\gamma<\gamma_{esc}$, will escape the emission site before contributing to the low energy flux. 
For $\gamma_{esc}<\gamma_m$ the spectrum below $\nu(\gamma_{esc})$ will be a regular slow cooling spectrum of $\nu^{-2/3}$.
Between $\nu(\gamma_{esc})$ and $\nu(\gamma_m)$ the spectrum will be fast cooling: $\nu^{-3/2}$.

If the escape is dominated by Bohm diffusion, then $t_{s} (\gamma) \propto \gamma^{-1}$, which is the marginal value of $a$.
In this case the ratio of the escape time to the cooling time is the same for all electron energies.
However, as all electrons are assumed to have initially been accelerated to above $\gamma_m$, electrons radiate first at $\nu\geq\nu_m$
and only later at lower frequencies. This means that for the marginal value of $a$, if $t_{s} (\gamma) \approx t_{c,s}(\gamma)$
(for $a=1$ if this ever equality holds, it does so for any value of $\gamma$),
then there would be a reduction of the low energy flux but not of the flux above $\nu_m$, as required by observations.
If $t_{s} (\gamma) > t_{c,s}(\gamma)$ the electrons don't escape and one gets the initial (non-escaping spectrum).
If $t_{s} (\gamma) \ll t_{c,s}(\gamma)$ the electrons cool down significantly and over produce optical and x-ray before escaping.

A typical electron has to spend a fraction $\sim t_{c,s}(\gamma_m)/t_p$ of the pulse duration within the radiation (high magnetic field) sites.
This implies that the overall fraction of the emitting volume occupied by the radiation sites, $x_V$ is of the same order.
To avoid significant cooling of the electrons outside the emission sites, the electrons' energy losses outside these sites
should be small compared with the energy loss within the sites:
\begin{equation}
\label{condition}
 \frac{P_{out} t_p'}{P_s t_{s}'} = \frac{B_{out}'^2 t_p'}{B_s'^2 t_{s} '}=\eta_B^{-2} \frac{t_p}{t_{c,s}(\gamma_m) }<1\rightarrow \eta_B^2 x_V>1 , 
\end{equation}
where $P \propto B^2$ is the synchrotron radiated power and we have used $t_{s} \approx t_{c,s}(\gamma_m)$ (from Eq. \ref{defescape}).
The condition in Eq.\ref{condition} is equivalent to requiring that the total magnetic energy is dominated by the emission sites.
Using the above estimate for $x_V$, the corresponding limit on $\eta_B$ is:
\begin{equation}
 \eta_B>10^9\G_{2.5}^{-8}(k+1)^{9/2}k^{-3/2}E_{tot,52.5}^{3/2} t_{p,-0.3}^{-5/2},
\end{equation}
Therefore the magnetic field in the sites must be huge compared to the background field.
In order for this configuration to be stable over a dynamical time of the system,
the magnetic energy density in the sites ($\eta_B^2$) must be compensated by a comparable particle pressure in the background.
The total energy will then be dominated by the particles and not the magnetic fields, contrary to our assumptions.
Note however, that such a model may still be relevant for resolving the low energy spectral index problem in a situation where the energy is not dominated by the magnetic field. 

\subsection{The confinement model}
An additional possibility, motivated by reconnection models, is that the magnetic field is dissipated in small regions within the magnetic background.
One may therefore have a situation in which, on average, $\epsilon_B \sim 1$, but electrons are accelerated and radiate within small
regions with much lower magnetic field. If the electrons remain confined in these regions, they cool less efficiently  and $\nu_c$ 
may increase, resulting in a spectrum  consistent with optical and x-ray observations.

In order to obtain values of $\nu_c\approx \nu_m$, we require that within the low-magnetization regions $\epsilon_B$ should be strongly suppressed:
\begin{equation}
\epsilon_{B}=2\times 10^{-6} \G_{2.5}^{16/3}(k+1)^{-3} k E_{tot,52.5}^{-1} t_{p,-0.3}^{2/3}.
\end{equation} 
Given the strong dependence on $\G$, it is not immediately clear what physically sets $\epsilon_B$ to this value in the low-magnetization sites.
Another implication of this configuration, is that the required typical $\gamma_e$ of the electrons becomes larger compared with the one zone model
(since the magnetic field was decreased, and we wish to keep $h\nu_m\approx 300$KeV):
\begin{equation}
\label{gmconf}
\gamma_m=10^5 \G_{2.5}^{-1/3} t_{p,-0.3}^{1/3}.
\end{equation}

For confinement,  the sizes of these regions
(in the co-moving frame) should be larger than the electrons' Larmour radius:
\begin{equation}
 r_L'=4\times 10^7 t_{p,-0.3}\mbox{cm}.
\end{equation}
If the geometry of the regions is such that particle trajectories naturally lead them to the boundaries, then since the magnetic fields are stronger in the background,
(by $\sim10^3$) the particles are better confined and the effective Larmour radius may become smaller by the same factor.
These radii should be compared to the plasma skin depth, which is the typical size on which plasma instabilities, such as reconnection
x-points, may develop:
\begin{equation}
\lambda'=8\times 10^2 \G_{2.5}^{2.8}(k+1)^{-1.5} k^{1/2} E_{tot,52.5}^{-1/2} t_{p,-0.3}^{5/3}\mbox{cm}
\end{equation}
It is obvious that $\lambda'\ll r_L'$ (even compared to the Larmour radii of the electrons on the background field) and therefore the particles are typically
not confined to the low magnetization sites.
Particles will therefore travel between x-points and "magnetic islands" as observed in numerical simulations \citep{Sironi(2014)}
and as a result will emit significant amounts of their energy in the high magnetic field region.

Finally, in the case of reconnection x-points, given the small skin depth, the life time of the low field regions is likely of the order of:
\begin{equation}
 \lambda'/c=3\times 10^{-8} \G_{2.5}^{2.8}(k+1)^{-1.5} k^{1/2} E_{tot,52.5}^{-1/2} t_{p,-0.3}^{5/3}\mbox{sec}
\end{equation}
much shorter than the dynamical time. Therefore, particles will eventually reside in high field regions in this scenario.

Thus, while it is not impossible, in principle, to resolve all these issues, these considerations demonstrate the limitations of inhomgenous magnetic field configurations in the presence of a background of Poynting flux dominated regions.

\section{Conclusions}
\label{conclusions}
Magnetically dominated jets need to dissipate some fraction of their energy to produce the observed radiation.
However, it is unclear where the dissipation takes place and how efficient is it.
In order to explore this issue we have explored the conditions within magnetically dominated 
emission regions, assuming that all the dissipation takes place right there.
We note that even if there is significant dissipation before the emitting region, the limits on the cooling time we find in the paper (discussed below) may be viewed as limits on the allowed dissipation rate.
Slower dissipation rates will effectively result in a magnetically dominated emission region, subject to the constraints found in this paper.
Alternatively the dissipation should not lead directly to particle acceleration and this will provide a ``buffer" during which the magnetic field decreases
but rapid electrons are not available to emit synchrotron.

We considered a general emission mechanism, assuming only that relativistic electrons are involved and focusing
on the inevitable synchrotron signature of these relativistic electrons in the presence of the strong magnetic fields.
We find that for $\Gamma\lesssim600$ relativistic electrons that are accelerated on a time scale shorter than the dynamical time scale,
will cool rapidly by synchrotron emission.
Therefore, any radiation mechanism proposed to explain the sub-MeV peak that relies on relativistic electrons should be faster than synchrotron. 
As such a mechanism is not known, synchrotron is the main emission mechanism in magnetically dominated systems. 
If synchrotron produces the prompt sub-MeV emission, the electrons cool extremely rapidly. 
For example, within an outflow with a typical Lorentz factor $\G=300$, the cooling time associated with the gamma ray emitting process, is at least one order of magnitude shorter than the
pulse time scale. The typical cooling frequency will be below the X-rays and in many cases even below optical.
This implies that synchrotron emission from these fast cooling electrons will be far above observational limits in the X-ray and in the optical.
This problem is related, but not directly similar, to the low energy spectral index of the prompt
$\gamma$ emission which is much harder than the one predicted by fast cooling.

We consider three possible ways to overcome the excessive emission at low frequencies. The electrons can be accelerated
before they cool down, they can escape rapidly from the large magnetic field zone or they may remain confined within sub-regions
with low magnetic field where they are accelerated and then radiate less efficiently.
The acceleration can be continuous \cite{Ghisellini(1999),Kumar(2008)}. In this case the electrons are kept more or less at the same Lorentz factor
during the whole emission episode posing a problem for production of the higher part of the energy spectrum. Alternatively the acceleration can be sporadic and the electrons are accelerated within
``re-acceleration sites" (this is the situation in e.g. the ICMART model \cite{Zhang(2011)}). The acceleration in these
``re-acceleration sites" should be modest, by  a factor less than three in energy. This implies that this scenario requires two 
different acceleration process. First an initial acceleration that brings the electrons to $\gamma_m \sim 5000$ and then 
a second re-acceleration process that accelerates the electrons mildly. An additional constraint  is that all 
electrons present must be accelerated in the first acceleration phase, otherwise slow electrons would be accelerated to 
mild energies within the ``re-acceleration sites" and will produce spurious low-energy synchrotron emission.
Due to the re-acceleration only very few electrons are needed to produce the observed flux. Hence, only very few protons must be present in the outflow.
Consequently all acceleration models require a very small baryonic loading (an extremely large $\sigma$).

An alternative scenario is one in which the magnetic field is inhomogeneous and the electrons cool efficiently only 
in ``emission sites" in which the magnetic field is stronger. However, this model requires that the magnetic field within
the emission sites should be significantly larger than the field in the outer regions.
This requires an external source of pressure that confines the magnetic field in the emission sites and it is incompatible
with the situation which we consider in which the outflow is Poynting flux dominated. This might be applicable, however, 
in other situations in which the magnetic field is weaker.

Finally, motivated by reconnection models, electrons could accelerate and radiate in small sub-volumes of the emitting region with lower magnetic fields, where their
cooling is less efficient. However, this model requires some extreme conditions. The magnetic fields within these regions should be very small
(typical values of $\epsilon_B \sim 10^{-6}$) leading to larger Larmour radii compared with the skin depth. Moreover, given the small skin depth, the life-time
of the low field regions is very likely to be much shorter than the dynamical time scale, and it is not clear that electrons can be confined sufficiently long in these low magnetic field regimes.

To summarize, we find that very specific conditions have to be satisfied if the emitting region is Poynting flux dominated. To be viable
any model in which the emission is in such a region must demonstrate how these conditions are reached. 
To our knowledge there is no known physical model that indeed demonstrates how these conditions can be met. 
The simplest possible resolution of this conundrum is that the outflow is not Ponynting flux dominated in the emission region.
This implies that Poynting flux dominated jets dissipate their magnetic energy before the emission zone.
Interestingly very different considerations, based on examination of propagations of relativistic
MHD jets within stellar envelope, have led  recently \cite{Bromberg+14} to reach a similar conclusion, namely that
Poynting flux dominated jets must dissipate deep in the interior of Collapsars in order to be compatible with observations.

We thank Jonathan Granot, Rodolfo Barniol Duran and Daniel Kagan for helpful discussions.
This research was supported by the ERC advanced research grant ``GRBs'',
by the  I-CORE (grant No 1829/12) and  by ISF-NSFC research grant.

\end{document}